\def\eqn#1{\eqno(#1)}
\def\mi{\medskip\noindent}
\def\bom{\mbox{\boldmath$\omega$\unboldmath}}
\def\xfg#1{fig.~\ref{#1}}
\documentstyle[12pt,epsfig]{article}
\begin{document}
\oddsidemargin 1mm
{\center
{\large\bf Generation of a symmetric magnetic field\\
by thermal convection in a plane rotating layer}

\mi
Vladislav Zheligovsky\footnote{E-mail: vlad@mitp.ru}

\mi
International Institute of Earthquake Prediction Theory\\
and Mathematical Geophysics\\
84/32 Profsoyuznaya St., 117997 Moscow, Russian Federation\\

\mi
Observatoire de la C\^ote d'Azur, CNRS\\
U.M.R. 6529, BP 4229, 06304 Nice Cedex 4, France\\

}

\medskip We investigate numerically magnetic field generation by thermal
convection with square periodicity cells in a rotating horizontal layer of
electrically-conducting fluid with stress-free electrically perfectly conducting
boundaries for Rayleigh numbers in the interval $5100\le R\le 5800$. Dynamos
of three kinds, apparently not encountered before, are presented: 1) Steady and
time-periodic regimes, where the flow and magnetic field are symmetric about
a vertical axis. In regimes with this symmetry, the global $\alpha-$effect is
insignificant, and the complex structure of the system of amplitude equations
controlling weakly nonlinear stability of the system to perturbations with
large spatial and temporal scales suggests that the perturbations are likely
to exhibit uncommon complex patterns of behaviour, to be studied
in the future work. 2) Periodic in time regimes, where magnetic field is always
concentrated in the interior of the convective layer, in contrast to the
behaviour first observed by St Pierre (1993) and often perceived as generic for
electrically infinitely conducting boundaries. 3) A dynamo exhibiting chaotic
behaviour of heteroclinic nature, where a sample trajectory enjoys excursions
between a periodic magnetohydrodynamic regime and rolls. The rolls are
amagnetic, but generate magnetic field kinematically. As a result, magnetic
energy falls off almost to zero, while the rolls are approached.


\section{Introduction}

Nonlinear regimes of magnetic field generation by Rayleigh-B\'enard convection
in elect\-rically conducting fluid in a plane layer have not yet been studied in
sufficient detail beyond the first bifurcations of the trivial steady state.
The only notable exception is the investigation by Podvigina \cite{P06c,P08b},
who traced the sequence of bifurcations and attractors for some parameter
values on increasing Rayleigh number in the absence of magnetic field, and
examined the ability of the convective flows to generate magnetic field in
kinematic and nonlinear regimes. A reasonably complete catalogue of convective
hydromagnetic (CHM) attractors in various regions of the parameter space,
presenting their symmetries and other essential properties, remains so far
unavailable. From this prospective, numerous studies performed at present,
where MHD or CHM simulations are carried out for extreme parameter values, are
of little interest: any emerging structural order in the regimes is ultimately
destroyed by chaos in the disguise of turbulence.

The goal of the present study is to look at a series of CHM
regimes not far from the onset of convection. We present dynamos of three
kinds, to the best of our knowledge not detected before:\\
$\bullet$ steady and periodic CHM regimes, symmetric about the vertical axis;\\
$\bullet$ dynamos with magnetic field concentrating in the middle of
the layer with perfectly conducting boundaries;\\
$\bullet$ a chaotic CHM regime where magnetic field occasionally switches off
almost entirely.\\
In the remaining part of the introduction we discuss in what respect
in our opinion such dynamos are remarkable.

\subsection{The symmetry about a vertical axis and parity invariance}

Our main interest lies in convective regimes of magnetic field generation,
which are parity-invariant or symmetric about a vertical axis, and stable to
perturbations of the same periodicity. Laminar dynamos
possessing various symmetries are worth to be identified, for instance,
because symmetries play an important r\^ole in the analysis of
dynamical systems and discovery of symmetric regimes can sometimes serve as
a basis for a subsequent mathematical investigation providing an insight
into different mechanisms of magnetic field generation. In this subsection we
discuss the fundamental reasons, coercing us to focus on CHM regimes with
the symmetries mentioned in the title. These reasons are related to stability
properties of CHM regimes with respect to large-scale perturbations.

Let us remind
the definitions of these symmetries. A CHM regime is parity-invariant, if
$\bf v$ and $\bf h$ are parity-invariant, and $\theta$ is parity-antiinvariant.
A vector field $\bf f$ and a scalar field $f$ are parity-invariant, if
$${\bf f}(-{\bf x},t)=-{\bf f}({\bf x},t);\qquad f(-{\bf x},t)=f({\bf x},t),$$
respectively; they are parity-antiinvariant, if in these equations
the signs of the right-hand sides are reversed. (The centre of symmetry
must lie on the mid-plane; without any loss of generality we assume
that the origin coincides with the centre of symmetry.) A CHM regime is
symmetric about the vertical axis, if all the three fields
$\bf v$, $\bf h$ and $\theta$ are. A vector field $\bf f$ and a scalar field
$f$ are symmetric about a vertical axis, if (again, without any loss
of generality assuming that the origin lies on the axis of symmetry)
$$f_1(-x_1,-x_2,x_3,t)=-f_1(x_1,x_2,x_3,t),$$
$$f_2(-x_1,-x_2,x_3,t)=-f_2(x_1,x_2,x_3,t),$$
$$f_3(-x_1,-x_2,x_3,t)=f_3(x_1,x_2,x_3,t),$$
and
$$f(-x_1,-x_2,x_3)=f(x_1,x_2,x_3),$$
respectively; they are antisymmetric, if in these equations the signs
of the right-hand sides are reversed. The two symmetries are compatible
with the equations (1)-(4) and with the boundary conditions (5)-(7)
considered in the present study (see the next section). For periodic regimes,
the symmetries can also involve a time shift by a half of the period.

In this and the following paragraph we summarise the findings of Zheligovsky
\cite{Zh08,Zh09}, who considered weakly nonlinear stability of short-scale CHM
regimes to large-scale perturbations (involving scales that are much larger than
the size of the periodicity cell of the perturbed CHM regime). Generically,
large-scale stability is controlled by the (global) combined kinetic and
magnetic $\alpha-$effect. Equations governing the evolution of weakly
nonlinear large-scale perturbations (more precisely, of the leading terms
in expansion in power series in the scale ratio $\varepsilon$ of averaged
in small scales amplitudes of neutral small-scale
stability modes, comprising the perturbation) are linear; thus,
they lack any inherent mechanism for saturation of perturbations. Moreover,
they involve no other terms but the slow time derivative and the operator
of the combined $\alpha-$effect, and the spectrum of the latter is symmetric
about the imaginary axis. Consequently, the equations generically
predict a superexponential growth of the perturbations resulting in an ultimate
destruction of the perturbed regimes on the time scales O($\varepsilon^{-1}$).

However, if any of the two symmetries under consideration are present,
the $\alpha-$effect is insignificant. Apparently (see \cite{CH06,HC08}; note
that different quantities are used {\it ibid.} and in \cite{Zh08,Zh09} to
measure the strength of the $\alpha-$effect -- they are based on the physical
common sense in the first case, and ensue from the multiscale asymptotic
analysis without a recourse to any empirical relation in the second case),
magnetic $\alpha-$effect can also disappear in highly turbulent regimes, but
simulations do not demonstrate this conclusively. In the absence of significant
combined $\alpha-$effect weakly nonlinear large-scale perturbations are
controlled by a system of nonlinear PDE's for amplitudes of neutral small-scale
stability modes. The system
is mixed: equations for the mean magnetic field are evolutionary, the rest ones
are not; it can involve cubic nonlinearity (see \cite{Zh09}). In contrast
to the case of the presence of significant $\alpha-$effect, a priori the
equations do not rule out saturation of perturbations at finite energy levels.
Instability to large-scale perturbations, if present, develops in MHD systems
with such symmetries on much larger time scales, O($\varepsilon^{-2}$).
Since in the absence of significant $\alpha-$effect weakly nonlinear
large-scale perturbations are controlled by a highly complex system of
nonlinear PDE's, the perturbations are likely to exhibit a variety of uncommon
complex patterns of behaviour.

One can look at the same problem at a different angle.
Many convective hydrodynamic and hydromagnetic regimes are essentially
non-space-periodic. The well-known examples include the K\"uppers-Lortz
\cite{KL69,CB79} and the small-angle
(see \cite{P08a,P09} and references therein) instabilities. Consequently,
simulations of convective hydrodynamic and MHD regimes in periodic boxes are
not physically sound. Moreover, convection in a horizontally unconstrained
domain can ``choose'' the preferred wave lengths -- e.g. on the onset of
convective motion the width of convective rolls can be uniquely determined
\cite{Chan}. One may try to overcome this difficulty in computations
by largely expanding the box in horizontal directions. The preferred wave
lengths of the convective system, which are a priori unknown, are approximated
the better, the larger is the box; however, the fundamental deficiency of this
approach (followed, for instance, in \cite{CH06,HC08}) -- in that the set
of wave lengths involved in computations is uniformly discrete -- remains
invincible. Perhaps, employment of asymptotic multiscale solutions in the style
of \cite{G94,Zh08,Zh09} is at present the only alternative method,
not suffering from this inadequacy, although it has its own limitations.

We are hence in need of examples of CHM regimes possessing any of the above
mentioned symmetries to carry out, in the future work, case studies of their
large-scale perturbations. Podvigina \cite{P06c,P08b} reported instances of
hydrodynamic steady and time-periodic convective regimes, that were symmetric
about a vertical axis and stable to short-scale perturbations (hence
non-chaotic) of the same period, as the perturbed short-scale CHM regime;
however, they failed to generate magnetic field with the same symmetry.
We have found steady and
time-periodic CHM regimes, which are symmetric about a vertical axis.

\subsection{Concentration of magnetic field in the middle of a layer\newline
with perfectly conducting boundaries}

We study convection in a plane rotating layer with stress-free perfectly
conducting boundaries held at constant temperatures.
Magnetic field generation by thermal convection with square periodicity cells
in the nonlinear regime was explored for these boundary conditions in
\cite{MP89,StP,Ma99,DS02,SH04,CH06,HC08,P06c,P08b}. Simulations
\cite{MP89,CH06,HC08} were performed for much higher Rayleigh numbers (of
the order of $10^5-10^6$) than the critical ones, resulting in the onset of
a turbulence-like behaviour of solutions. In the remaining studies cited above
the Rayleigh number did not exceed a moderate multiple of the critical value
and convective flows exhibited laminar behaviour, although instances of chaotic
behaviour of MHD systems -- precursors to turbulence -- were also obtained.
The case of fast rotation was considered in \cite{StP,SH04}; the Rayleigh
number was kept below the critical value for the onset of non-magnetic
convection in \cite{StP}. Magnetic field showed
a variety of patterns of behaviour.

In the convective dynamos
\cite{StP,Ma99} magnetic field concentrated near the boundaries in
sublayers of the width of a quarter of the layer width. Magnetic energy attained
the maximum on the boundary in the run labeled N6 in \cite{MP89}
(see a snapshot of magnetic field on fig.~6 {\it ibid.}; no detailed information is
presented about the spatial distribution of magnetic field in other runs).
St Pierre \cite{StP} suggested a physical mechanism for explanation of
concentration of magnetic field near infinitely electrically conducting
boundaries: if the fluid motion in the vertical direction is not
inhibited, magnetic field is advected to a boundary and remains ``locked''
there, unable to exit from the fluid layer outside the boundary acting
as a screen. Thus, these boundary conditions can be beneficial for magnetic
field generation. (In dynamo simulations \cite{TH00}
in a plane convective layer and \cite{SB05} in a rotating
spherical fluid shell, the efficiency of dynamo was not significantly
affected by the boundary conditions for magnetic field, which
were varied only on the inner core boundary in \cite{SB05}.) In the convective
dynamos \cite{P06c,P08b} in a non-rotating layer, strong magnetic field
usually formed ropes near the boundaries, however, Podvigina \cite{P08b} found
chaotic regimes, where magnetic field was residing part of the time
near the boundaries and part of the time inside the layer. Magnetic field
computed in \cite{DS02} had a form of cigar-like structures
of vertical orientation, mostly located inside the layer, with small blobs
of strong field near the horizontal boundaries. In the convective dynamo
computations \cite{SH04} for a much faster than in our simulations rotation,
magnetic field was organised into irregular structures with magnetic energy
most time residing inside the fluid layer.

We have encountered CHM time-periodic regimes, in which, in contrast with all
the simulations mentioned above, the generated magnetic field {\it always}
remains concentrated in the interior of the convective layer.

\subsection{A heteroclinic CHM regime\newline
with occasional disappearance of magnetic field}

We have found a chaotic CHM regime of a heteroclinic nature, where a trajectory
jumps between a CHM periodic orbit and amagnetic steady rolls. Both, the orbit
and the rolls, are slightly unstable. The rolls are capable of kinematic
magnetic field generation, however, while the trajectory approaches the rolls
after the regime has deviated from the periodic oscillations, magnetic energy
decays almost to zero before the growing magnetic mode becomes substantial.

From the viewpoint of the theory of dynamical systems, nothing
is particularly surprising in such a behaviour: heteroclinic cycles between
a periodic orbit and two steady states were studied in \cite{Me89} and
were observed, for instance, in \cite{Me01} in a low-order dynamical
system, modelling excursions and reversals of the Earth's magnetic field.
To the best of our knowledge, such behaviour was never before found
in numerical simulations of ``real'' physical systems. It is notable, that
this CHM regime and regimes observed for some sets of parameter values
in the VKS (von K\'arm\'an Sodium) experiment \cite{Pint09} exhibit certain
similar features, such as intermittency with bursts and extinction
of magnetic field (see fig. 25 {\it ibid.}) and heteroclinic connections
of unstable MHD steady states (but the regimes of heteroclinic nature
apparently do not involve connections to periodic orbits).

\section{Numerical simulation}

Evolution of the Boussinesq CHM system satisfies the non-dimensional
Navier-Stokes, magnetic induction and heat transfer equations:
$${\partial{\bf v}\over\partial t}=P\nabla^2{\bf v}+{\bf v}\times\bom
-{\bf h}\times(\nabla\times{\bf h})
+P\tau{\bf v}\times{\bf e}_3+PR\theta{\bf e}_3-\nabla p,\eqn{1}$$
$${\partial{\bf h}\over\partial t}={P\over P_m}
\nabla^2{\bf h}+\nabla\times({\bf v}\times{\bf h}),\eqn{2}$$
$${\partial\theta\over\partial t}=\nabla^2\theta
-({\bf v}\cdot\nabla)\theta+v_3,\eqn{3}$$
and the solenoidality conditions
$$\nabla\cdot{\bf v}=0,\qquad\nabla\cdot{\bf h}=0.\eqn{4}$$
Here $\bf h$ and $\bom=\nabla\times\bf v$ denote velocity and vorticity,
respectively, of a flow of an electrically conducting fluid, $p$ modified
pressure, $\bf h$ magnetic field, $\theta=T-(x_3+L/2)$
the difference between the normalised temperature $T$ and the linear
temperature profile in the layer in the absence of fluid motion; $t$ is time,
$P$, $P_m$, $R$ and $\tau^2$ are kinematic Prandtl, magnetic Prandtl, Rayleigh
and Taylor numbers, respectively, $\tau/2$ is the normalised angular rotation
rate, ${\bf e}_k$ the unit vector along the axis $x_k$ of the Cartesian
coordinate system co-rotating with the layer; $L$ is the width of the layer.
The CHM system is free: no
external sources are present in equations (1)-(3). Hence, the regimes
are spatially and temporally invariant -- there is no explicit dependence of
any term in the equations either on the position in space, or on time.

We assume that the horizontal boundaries of the layer are stress-free:
$$\left.{\partial v_1\over\partial x_3}\right|_{x_3=\pm L/2}=
\left.{\partial v_2\over\partial x_3}\right|_{x_3=\pm L/2}=0,\quad
\left.v_3\right|_{x_3=\pm L/2}=0,\eqn{5}$$
infinitely electrically conducting:
$$\left.{\partial h_1\over\partial x_3}\right|_{x_3=\pm L/2}=
\left.{\partial h_2\over\partial x_3}\right|_{x_3=\pm L/2}=0,\quad
\left.h_3\right|_{x_3=\pm L/2}=0,\eqn{6}$$
and kept at fixed temperatures:
$$\theta\left|_{x_3=\pm L/2}=0.\right.\eqn{7}$$
These boundary conditions are the most convenient ones for simulation
of Boussinesq convection in the layer. They were used in all numerical dynamo
simulations cited in the Introduction, (5) was assumed in the investigation
\cite{P08a,P09} (also see references therein) of the hydrodynamic small-angle
instability, and stability regimes to large-scale perturbations were studied
in \cite{Zh08,Zh09} also for (5)-(7).

We have carried out a numerical search for examples of symmetric CHM
attractors with the same period $L_h$ along the horizontal coordinate axes
$x_1$ and $x_2$. To avoid turbulisation of the flow and symmetry breaking,
we have not used extreme parameter values like in some studies cited
in the Introduction. Simulations have been performed for $L=1$, $L_h=10/7$,
$P=1$, $\tau=91$ (i.e. the Taylor number is 8281) for $R$ increasing from
5100 to 5800. For these values, the critical Rayleigh number
for the onset of convective flow in the absence of magnetic field is
equal to $\sim$5514; for the periodicity box, which is $\sqrt2$ times
smaller in size and aligned with the diagonals of the computational periodicity
box, the critical Rayleigh number is $\sim$5072 (the two critical numbers have
been evaluated using a precise formula for $R$ derived in \cite{Chan} for
the boundary conditions assumed here).

As discussed in the Introduction, our prime goal is to find symmetric CHM
regimes. In the primary (hydrodynamic) bifurcation convective flows in the form
of rolls set in, which possess the desirable symmetry. Because of rotation,
they can (at least in principle) kinematically generate magnetic field, if
magnetic Prandtl number is high enough; if the dominant magnetic mode also has
the symmetry, then the CHM regime emerging in the secondary (magnetic)
bifurcation remains symmetric. Consequently, we have chosen to perform
simulations for a (relatively) high value $P_m=100$ (in contrast with $P_m<1$
used for numerical simulations of the Earth's dynamo, characterised
by the values $P_m\sim 10^{-5}$).

Unlike in \cite{P06c,P08b}, we have
not attempted to analyse the sequence of occurring bifurcations in detail, or
to locate them to a good accuracy, nor to identify all attractors of the CHM
system for any considered $R$.

\begin{table}[t]
Table. Regimes of thermal hydromagnetic convection.
$E^k_{\min},\ E^k_{\rm av},\ E^k_{\max}$: kinetic energy per periodicity cell,
minimal, average over a period and maximal, respectively;
$E^m_{\min},\ E^m_{\rm av},\ E^m_{\max}$: the same for magnetic energy.
SVA: symmetry about the vertical axis ($+$: $\bf v$ and $\bf h$ are both
symmetric; $-$: $\bf v$ is symmetric and $\bf h$ is antisymmetric;
0: $\bf v$ is symmetric, no magnetic field generation;
No: $\bf v$ and $\bf h$ do not possess the symmetry or antisymmetry).
P: parity related symmetries (see the text). Type: the type of regime
(SS: a steady state, DR: steady rolls aligned with the diagonal, with
$n_1\!=\!n_2$ and $n_3$ of the same parity in (8), HC: heteroclinic chaos,
a number: temporal frequency of a periodic orbit).

\medskip
\centerline{\begin{tabular}{|c|c|c|c|c|c|c|c|c|c|}\hline
$R$ & $E^k_{\min}$ & $E^k_{\rm av}$ & $E^k_{\max}$ & $E^m_{\min}$ & $E^m_{\rm av}$ & $E^m_{\max}$ & SVA & P & Type \\\hline
5100 & 1.0 & 1.0 & 1.0 & 0.012 & 0.012 & 0.012 & $+$ & $\sigma_1$ & SS\\\hline
5140 & 2.4 & 2.4 & 2.4 & 0.025 & 0.025 & 0.025 & $+$ & $\sigma_1$ & SS\\\hline
5160 & 4.9 & 5.3 & 5.8 & 1.30 & 1.46 & 1.71 & $-$ & & 0.18 \\\hline
5180 & 5.8 & 6.3 & 7.0 & 1.58 & 1.84 & 2.28 & $-$ & & 0.19 \\\hline
5190 & 6.3 & 6.9 & 7.7 & 1.75 & 2.10 & 2.68 & $-$ & & 0.19 \\\hline
5210 & 5.2 & 6.8 & 16.1 & 0.33 & 1.86 & 12.8 & $-$ & & HC \\\hline
5210 & 6.5 & 6.5 & 6.5 & 0 & 0 & 0 & 0 & & DR\\\hline
5220 & 6.9 & 6.9 & 6.9 & 0 & 0 & 0 & 0 & & DR\\\hline
5230 & 6.7 & 6.7 & 6.7 & 0 & 0 & 0 & 0 & & DR\\\hline
5240 & 7.9 & 7.9 & 7.9 & 0 & 0 & 0 & 0 & & DR\\\hline
5250 & 8.4 & 8.4 & 8.4 & 0 & 0 & 0 & 0 & & DR\\\hline
5260 & 8.8 & 8.8 & 8.8 & 0 & 0 & 0 & 0 & & DR\\\hline
5300 & 10.8 & 10.8 & 10.8 & 0 & 0 & 0 & 0 & & DR\\\hline
5350 & 9.8 & 10.0 & 10.4 & 0.30 & 0.55 & 0.95 & $+$ & $\sigma_1$ & 0.315 \\\hline
5400 & 11.5 & 12.0 & 13.0 & 0.26 & 0.75 & 1.85 & $+$ & $\sigma_1$ & 0.31 \\\hline
5500 & 15.1 & 16.0 & 19.0 & 0.27 & 1.13 & 3.71 & $+$ & $\sigma_1$ & 0.32 \\\hline
5550 & 16.9 & 18.0 & 22.9 & 0.25 & 1.15 & 4.68 & $+$ & $\sigma_1$ & 0.29 \\\hline
5600 & 20.3 & 22.3 & 25.5 & 1.56 & 2.30 & 3.77 & No & $\sigma_2$ & 0.66 \\\hline
5700 & 24.6 & 27.3 & 31.0 & 2.20 & 3.20 & 4.91 & $-$ & $\sigma_2$ & 0.76 \\\hline
5750 & 26.3 & 29.8 & 33.9 & 2.28 & 3.71 & 5.62 & No & $\sigma_2$ & 0.38 \\\hline
5800 & 26.1 & 33.0 & 39.7 & 0 & 3.70 & 27.12 & No & $\sigma_3$ & HC \\\hline
\end{tabular}}
\end{table}

For the boundary conditions under consideration, the fields are expanded
in the Fourier series
$${\bf v}=\sum_{{\bf n}:\ n_3\ge0}\left[
\begin{array}{c}
v_{{\bf n},1}\cos n_3\pi(x_3/L+1/2)\\
v_{{\bf n},2}\cos n_3\pi(x_3/L+1/2)\\
v_{{\bf n},3}\sin n_3\pi(x_3/L+1/2)
\end{array}\right]{\rm e}^{{2\pi i\over L_h}(n_1x_1+n_2x_2)},\eqn{8}$$
$${\bf h}=\sum_{{\bf n}:\ n_3\ge0}\left[
\begin{array}{c}
h_{{\bf n},1}\cos n_3\pi(x_3/L+1/2)\\
h_{{\bf n},2}\cos n_3\pi(x_3/L+1/2)\\
h_{{\bf n},3}\sin n_3\pi(x_3/L+1/2)
\end{array}\right]{\rm e}^{{2\pi i\over L_h}(n_1x_1+n_2x_2)},\eqn{9}$$
$$\theta=\sum_{{\bf n}:\ n_3>0}\theta_{\bf n}\sin n_3\pi(x_3/L+1/2)\,
{\rm e}^{{2\pi i\over L_h}(n_1x_1+n_2x_2)}.\eqn{10}$$
Solenoidality conditions (4) for the series (8) and (9) take the forms
$${2\pi in_1\over L_h}v_{{\bf n},1}
+{2\pi in_2\over L_h}v_{{\bf n},2}+{\pi n_3\over L}v_{{\bf n},3}=0;$$
$${2\pi in_1\over L_h}h_{{\bf n},1}
+{2\pi in_2\over L_h}h_{{\bf n},2}+{\pi n_3\over L}h_{{\bf n},3}=0.$$
Standard pseudospectral methods (see \cite{Ca88,Boyd,Pe02})
have been applied with the resolution of $64\times64\times32$ Fourier harmonics
(after dealiasing in computation of advective terms was performed). The energy
spectra of the computed attractors decay by several (at least 3-4 for magnetic
field, which has the least decreasing energy spectrum) orders of magnitude.
No significant discrepancies with the results of several test runs with the
$128\times128\times64$ resolution have been found.

\section{Results of computation}

Properties of the observed CHM attractors are summarised in the table.
In the interval $5210\le R\le 5300$ they degenerate into a system of
two-dimensional rolls aligned with the diagonal of the periodicity box,
in which the flow is independent from the position along the axes of
the rolls. Although the flows in the rolls has all the three spatial components
and the Zeldovich antidynamo theorem \cite{Zel} is not applicable, the rolls
do not generate magnetic field. In all other saturated regimes that have been
simulated, generation takes place.

For most attractors found in computations, the sums (8)-(10) turn out to be
comprised of terms, where for some $i,j=1,2,3$ the sum $n_i+n_j$ has a certain
parity (see the table). Suppose $v_{{\bf n},i}\ne 0$ only, if $n_i+n_j$ is even.
This is consistent with the action of the Lorentz force (quadratic in $\bf h$)
in the Navier-Stokes equation (1) only, if for all non-zero terms in (8) parity
of $n_i+n_j$ is the same -- either odd, or even. If a scalar or vector field is
represented by a sum of the form of (10) or (8), respectively, where $n_1+n_2$
is even, then the size of its elementary periodicity cell is $L_h/\sqrt2$, and
the cell sides are oriented along the diagonals of the computational
periodicity cell. In particular, for $R=5100$, 5140 and $5350\le R\le 5700$
$v_{{\bf n},i}\ne 0$ only, if $n_1+n_2$ is even, and $h_{{\bf n},i}\ne 0$ only,
if $n_1+n_2$ is odd. Thus, the periodicity cell of the flow is $\sqrt2$ times
less than the periodicity cell of magnetic field, i.e. a mild scale separation
is present in these CHM regimes. If coefficients of a field do not vanish
only, if $n_j+n_3$ is even ($j=1$ or 2), it has a symmetry, which is a
composition of translation by a half of period along the axis $x_j$ and
reflection about the mid-plane. The following three combinations of symmetries
of this kind have been encountered in computations:\\
$\sigma_1$: $v_{{\bf n},i}\ne 0$ only, if
${\rm mod}_2\,n_1={\rm mod}_2\,n_2={\rm mod}_2\,n_3$ and $h_{{\bf n},i}\ne 0$
only, if ${\rm mod}_2\,n_1\ne{\rm mod}_2\,n_2={\rm mod}_2\,n_3$;\\
$\sigma_2$: $v_{{\bf n},i}\ne 0$ only, if
${\rm mod}_2\,n_1={\rm mod}_2\,n_2={\rm mod}_2\,n_3$ and $h_{{\bf n},i}\ne 0$
only, if ${\rm mod}_2\,n_1={\rm mod}_2\,n_2\ne{\rm mod}_2\,n_3$;\\
$\sigma_3$: $v_{{\bf n},i}\ne 0$ only, if ${\rm mod}_2\,n_2={\rm mod}_2\,n_3$ and
$h_{{\bf n},i}\ne 0$ only, if ${\rm mod}_2\,n_2\ne{\rm mod}_2\,n_3$.
(We use here the standard notation ${\rm mod}_2\,n$ for parity of $n$: it is
0 for even $n$, and 1 for odd $n$.)

Except for $R=5600$ and 5800, the flows in attractors are symmetric about
the vertical axis (see the table); in agreement with this, magnetic field either
has the same symmetry, or it is antisymmetric. All the simulated CHM regimes,
in which all the three constituent fields possess this symmetry, exhibit
a simple time dependence: they are steady or periodic, and in addition
have the symmetries $\sigma_1$, in particular, implying scale separation.

\begin{figure}[p]
\centerline{\psfig{file=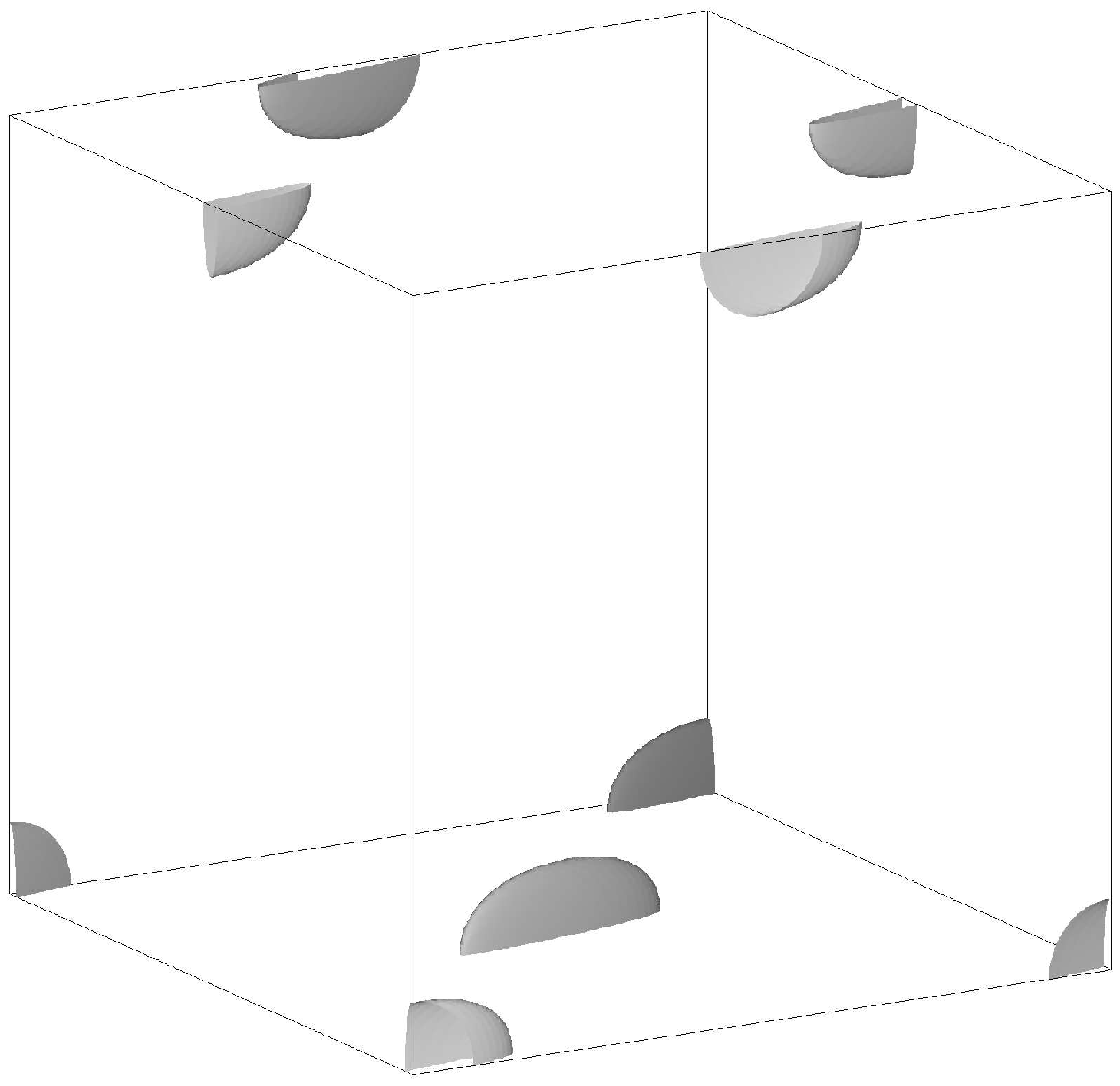,width=7cm,height=6cm,clip=}\hspace{1cm}
\psfig{file=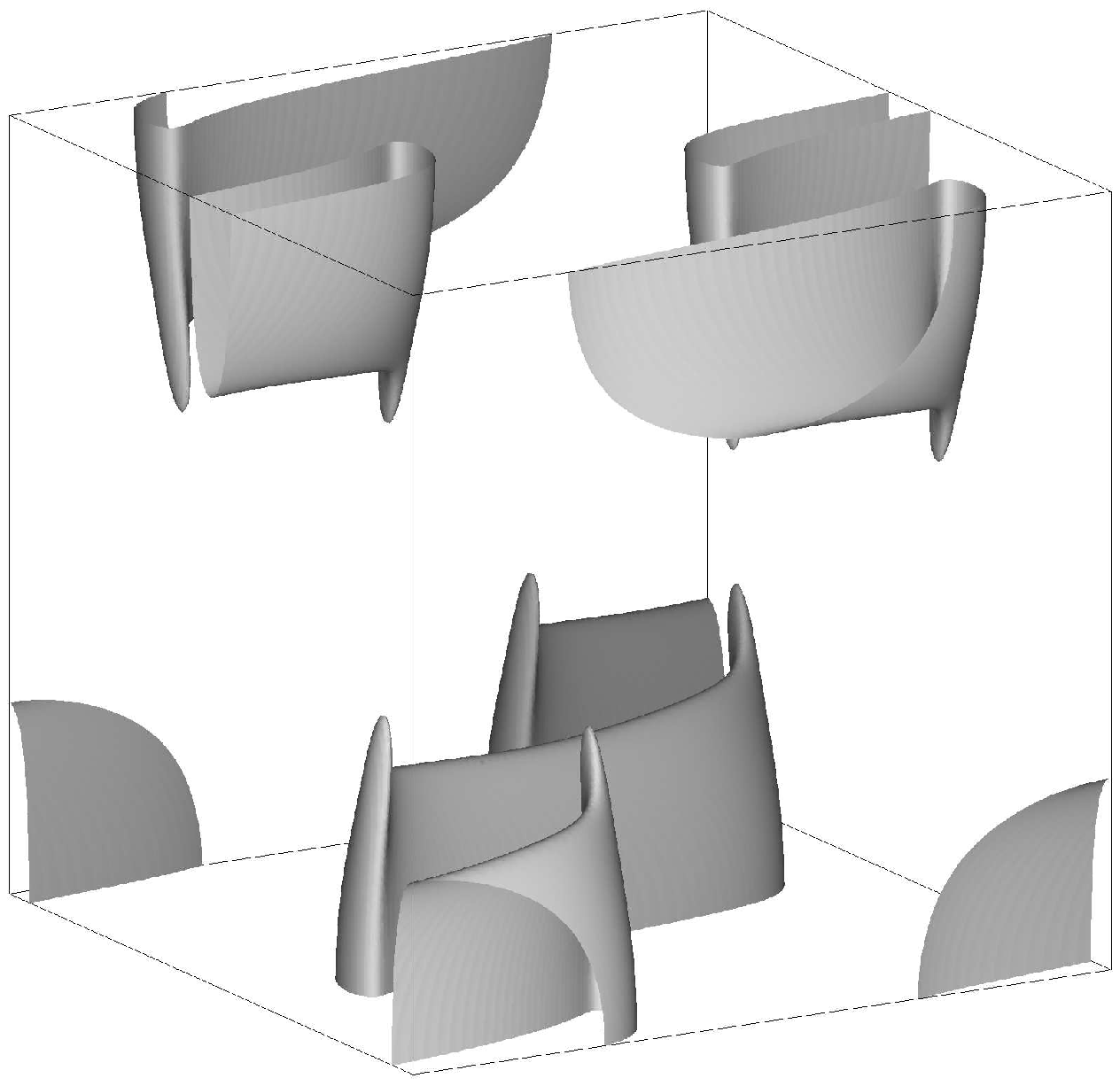,width=7cm,height=6cm,clip=}}

\vspace*{-7mm}
\hspace{4mm}(a)\hspace{76mm}(b)

\caption{Isosurfaces of magnetic energy density $|{\bf h}|^2$
at the levels of 2/3 (a) and 1/20 (b) of the maximum for the
CHM steady state for $R=5100$. One periodicity cell is shown.}
\label{f:b5100}

\bigskip
\centerline{\raisebox{1cm}{(a)}\ \ \psfig{file=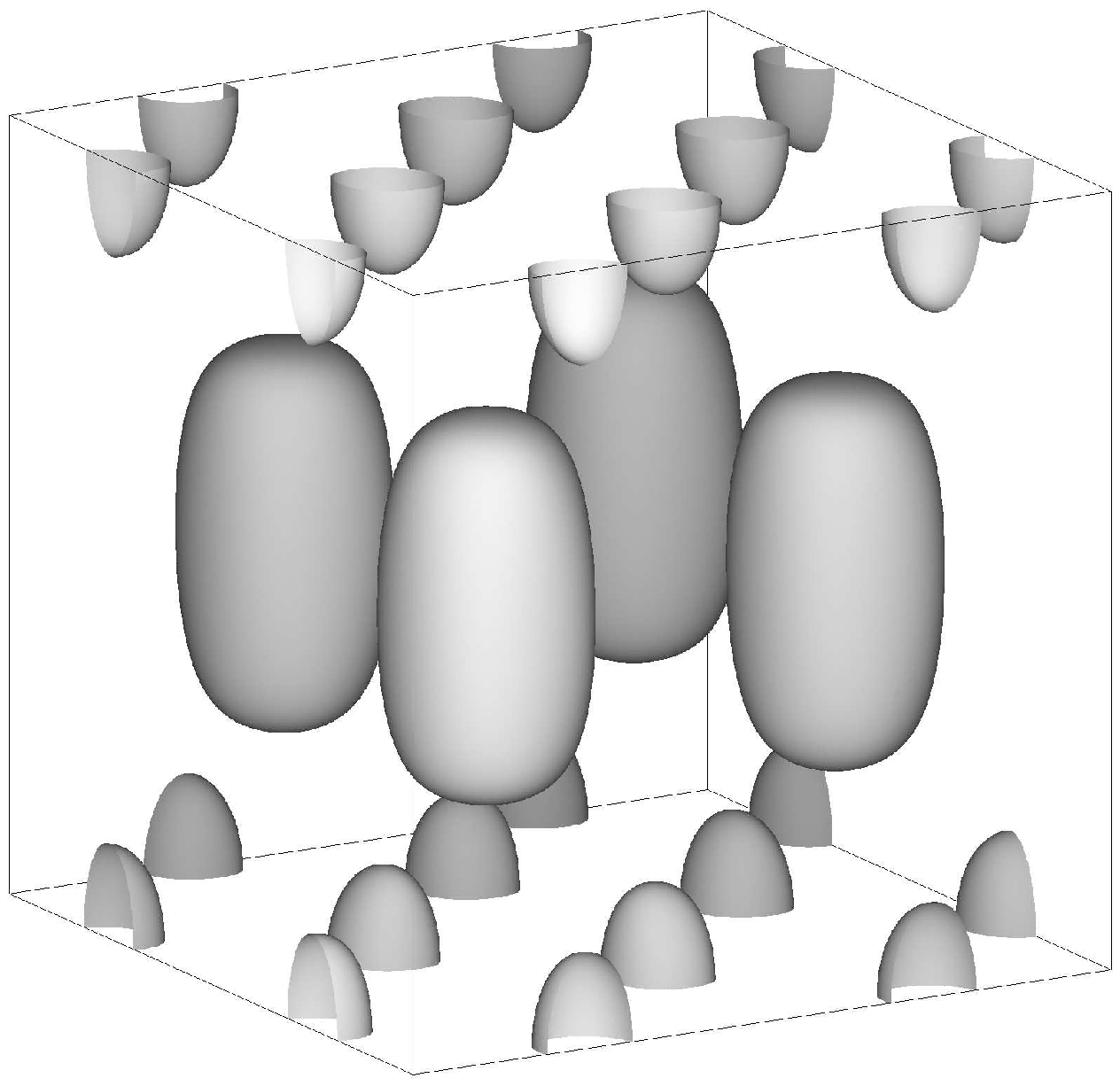,width=7cm,height=6cm,clip=}}
\vspace*{6mm}

\centerline{\raisebox{1cm}{(b)}\ \ \psfig{file=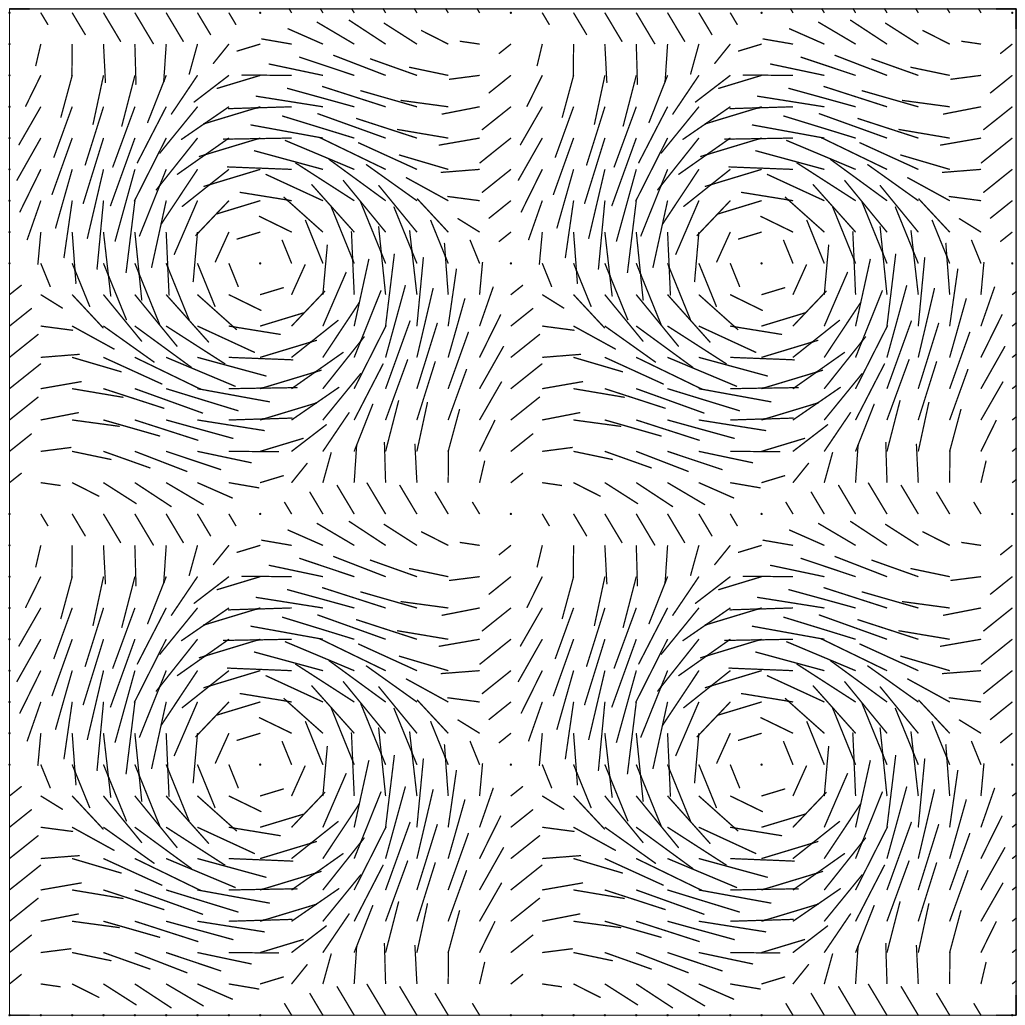,height=6cm}\hspace{15mm}
\raisebox{1cm}{(c)}\ \ \psfig{file=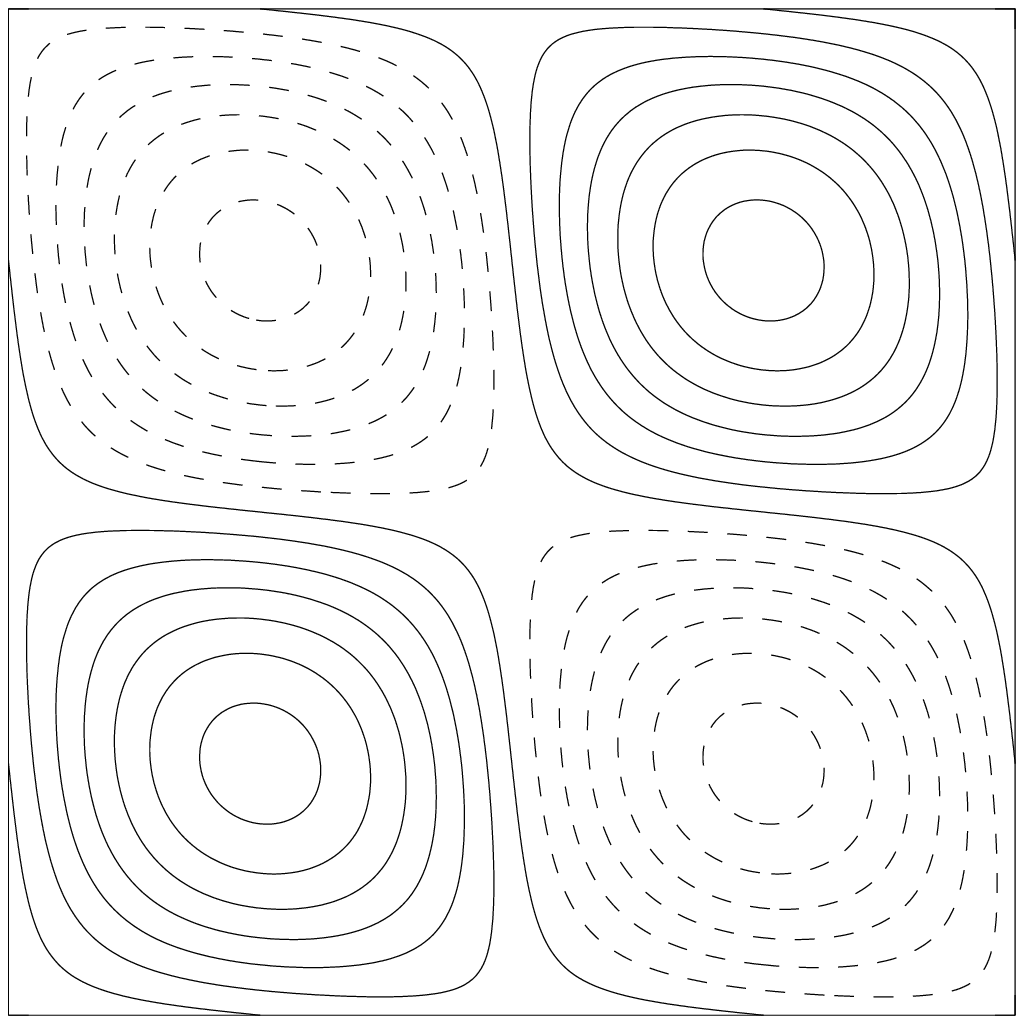,height=6cm}}

\caption{Isosurfaces of kinetic energy density $|{\bf v}|^2$ at the level
of a half of the maximum (a) and the flow on the mid-plane $x_3=0$: horizontal
components (b) of the velocity of the fluid and isolines of the vertical
component (c) of the velocity step 0.3 (zero and positive values: solid lines,
negative values: dashed lines) for the CHM steady state for $R=5100$.
One periodicity cell is shown.}
\label{f:v5100}\end{figure}

\begin{figure}[t]
\centerline{\psfig{file=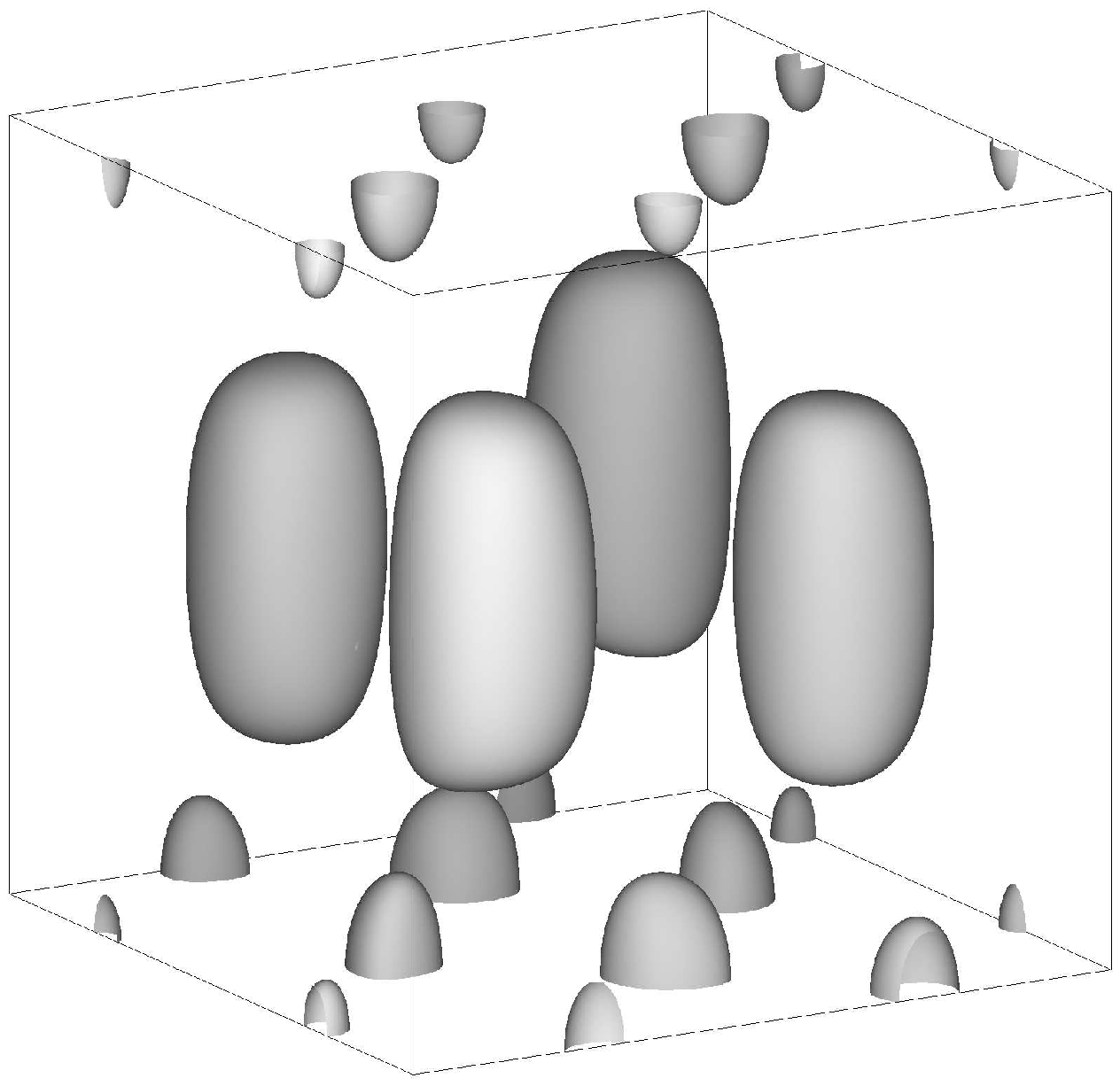,width=7cm,height=6cm,clip=}}
\caption{Isosurfaces of kinetic energy density $|{\bf v}|^2$ at the level
of a half of the maximum for the time-periodic CHM regime for $R=5180$.
One periodicity cell at a randomly chosen time is shown.}
\label{f:v5180}

\bigskip
\centerline{\psfig{file=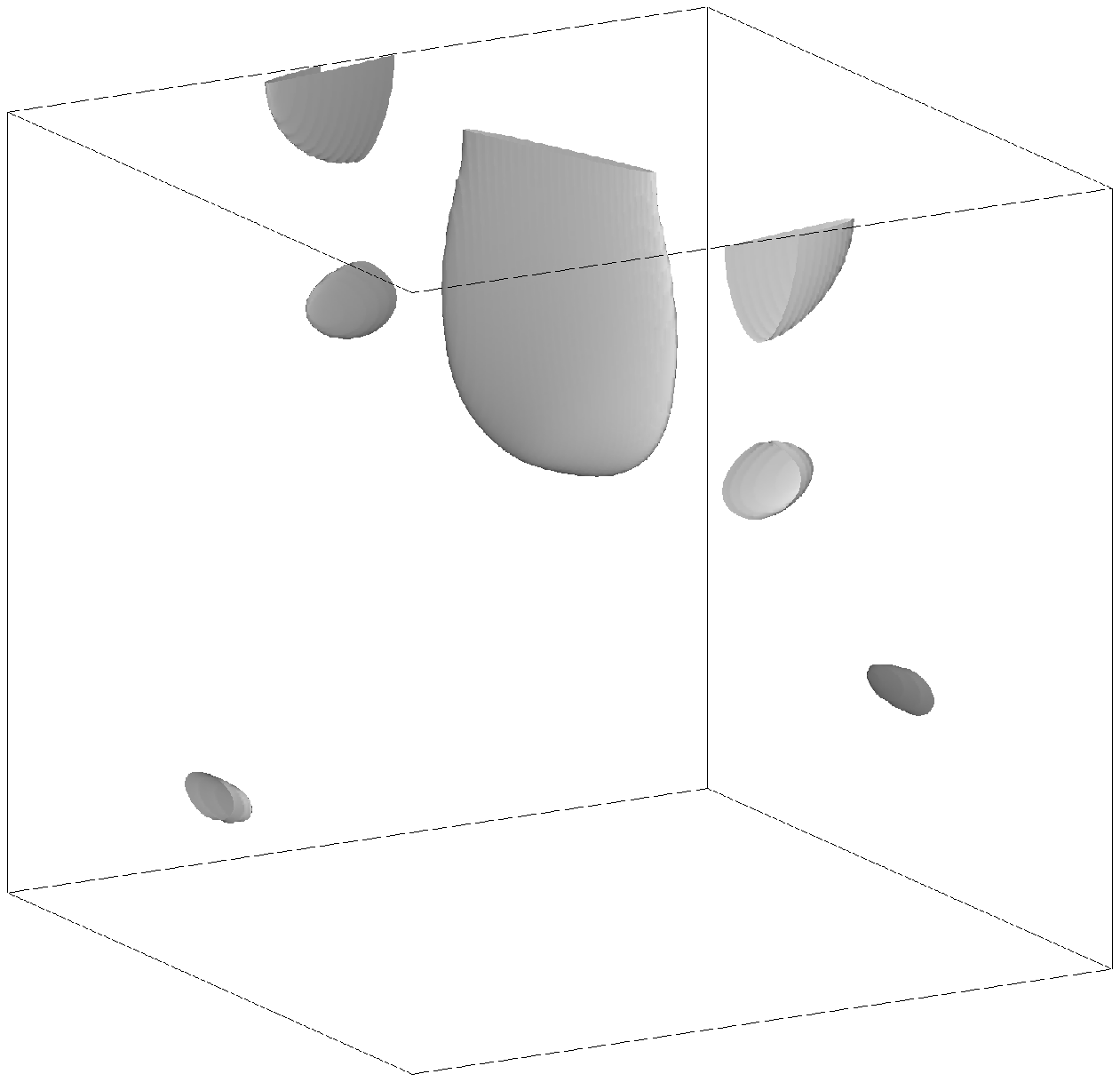,width=7cm,height=6cm,clip=}\hspace{1cm}
\psfig{file=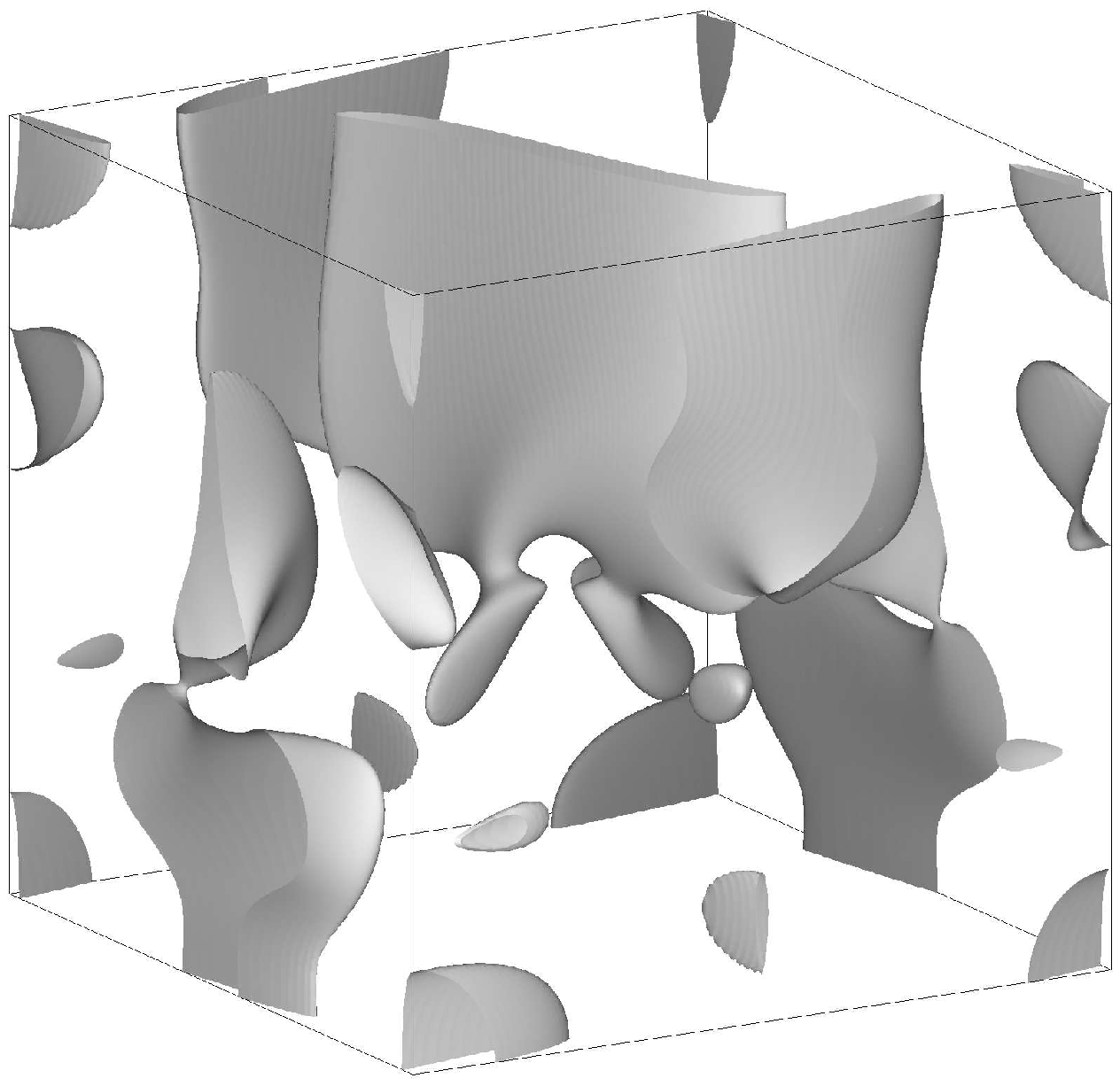,width=7cm,height=6cm,clip=}}

\vspace*{-7mm}
\hspace{4mm}(a)\hspace{76mm}(b)

\caption{Isosurfaces of magnetic energy density $|{\bf h}|^2$ at the levels
of 1/2 (a) and 1/10 (b) of the maximum for the time-periodic CHM regime for
$R=5180$. One periodicity cell is shown at the time, for which
\protect\xfg{f:v5180} is constructed.}
\label{f:b5180}\end{figure}

CHM regimes for $5100\le R\le 5140$ are steady states (see figs.~\ref{f:b5100}
and \ref{f:v5100}). As documented in \cite{StP}, magnetic field generated
by convective flows tends to concentrate near infinitely electrically
conducting boundaries of the layer. Magnetic flux ropes (\xfg{f:b5100}a)
generated by steady flows are associated with stagnation points of the flow,
which have one-dimensional unstable and two-dimensional stable manifolds
(asymptotic solutions of the magnetic induction equation describing magnetic
ropes in the vicinity of such stagnation points were presented in
\cite{Moff,Zh93,GZ94}); the ropes are
cut in halves along their axes by the boundaries of the layer. In the steady CHM
state for $R=5100$, stagnation points of the flow are located on
the midplane and on the upper and lower boundaries of the box of periodicity
shown on \xfg{f:v5100}. On each of the three horizontal planes they constitute
a square mesh of mesh size $L_h/\sqrt{8}$, oriented along the diagonals of
the periodicity cells. Only stagnation points at the centre and in the vertices
of the horizontal faces of the shown box of periodicity are associated with
the magnetic flux ropes. The flow can be described as a system of downward and
upward swirls (\xfg{f:v5100}) positioned in the chessboard order. Inward flows
advect magnetic field from the magnetic ropes near the boundaries towards
the middle of the layer with formation of magnetic flux sheets
between adjacent swirls (\xfg{f:b5100}b).

\begin{figure}[t]
\centerline{\psfig{file=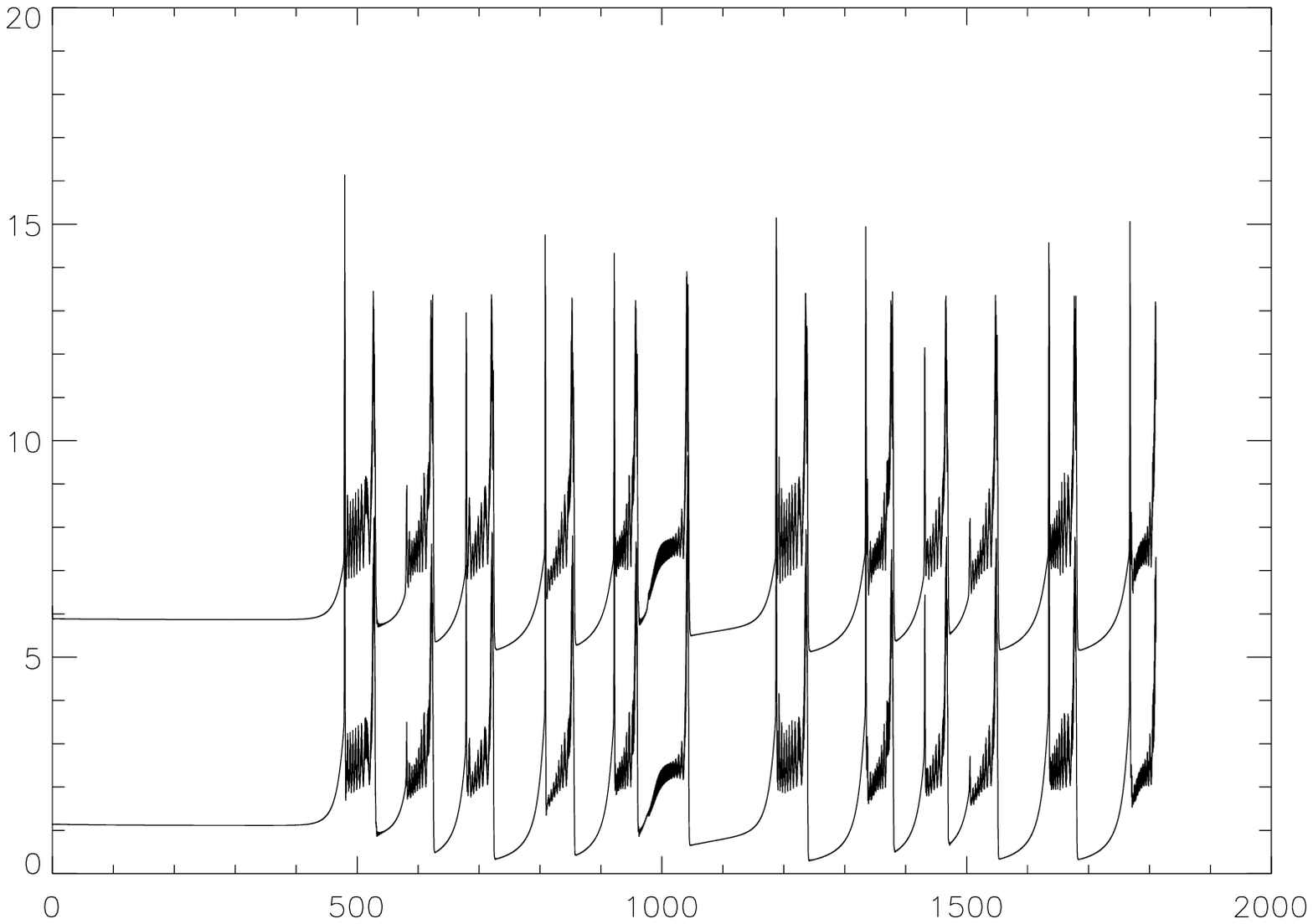,width=15cm,height=9cm,clip=}}
\caption{Total kinetic (upper curve) and magnetic (lower curve) energies
(vertical axis) as a function of time (horizontal axis) for the chaotic
CHM regime for $R=5210$.}
\label{f:e5210}\end{figure}

A branch of periodic orbits exists for $5160\le R\le 5190$. (For symmetry
reasons, this branch of periodic regimes does not emerge from the branch
of steady states.) The symmetry about the vertical axis of the flow and
temperature is preserved, but magnetic field is now antisymmetric (and
significantly stronger). Unlike for the CHM steady states
observed for smaller considered $R$, there is no parity selection of
wave numbers in the Fourier series (8)-(10) representing the solutions.
However, the structures of the flow and (to a lesser extent) magnetic field
resembles those for $R=5100$ (\xfg{f:v5180}, \ref{f:b5180}a). The main feature
of magnetic field is vertical magnetic flux sheets (\xfg{f:b5180}b), resembling
sheets developing in the vicinity of an unstable manifold of a stagnation point
of the flow, if the point has a two-dimensional unstable and one-dimensional
stable manifolds (asymptotic analysis of solutions of the magnetic induction
equation in this context was carried out by Childress and Soward 1985).

At $R=5210$ two CHM attractors are found in the system: amagnetic steady rolls
and an attractor exhibiting heteroclinic chaotic behaviour (\xfg{f:e5210}).
In the phases of oscillations of growing amplitude the system is
in the vicinity of the former periodic orbit. The sample trajectory
vigorously exponentially departures from it, but subsequently comes
to the phase of lower kinetic and magnetic energy levels, during which it is
in the vicinity of a mildly unstable steady state, near which integration
of the sample trajectory has begun. Duration of this phase can be quite short.
The phase of evolution near the steady state also finishes in an exponential
departure, with kinetic and magnetic energy surges of amplitude comparable
to that of energy surges during departures from the phase of evolution near
the periodic orbit. Sample trajectories which start in the vicinity of the same
steady state for higher $R$, are attracted to steady rolls in shorter times,
than the length of integration of the trajectory for $R=5210$; however, their
behaviour at the intermediate transitory stage resembles the chaotic behaviour
for $R=5210$, and thus from a conservative point of view it cannot be excluded,
that duration of the chaotic stage at $R=5210$ is also finite.

\begin{figure}[p]
\centerline{\psfig{file=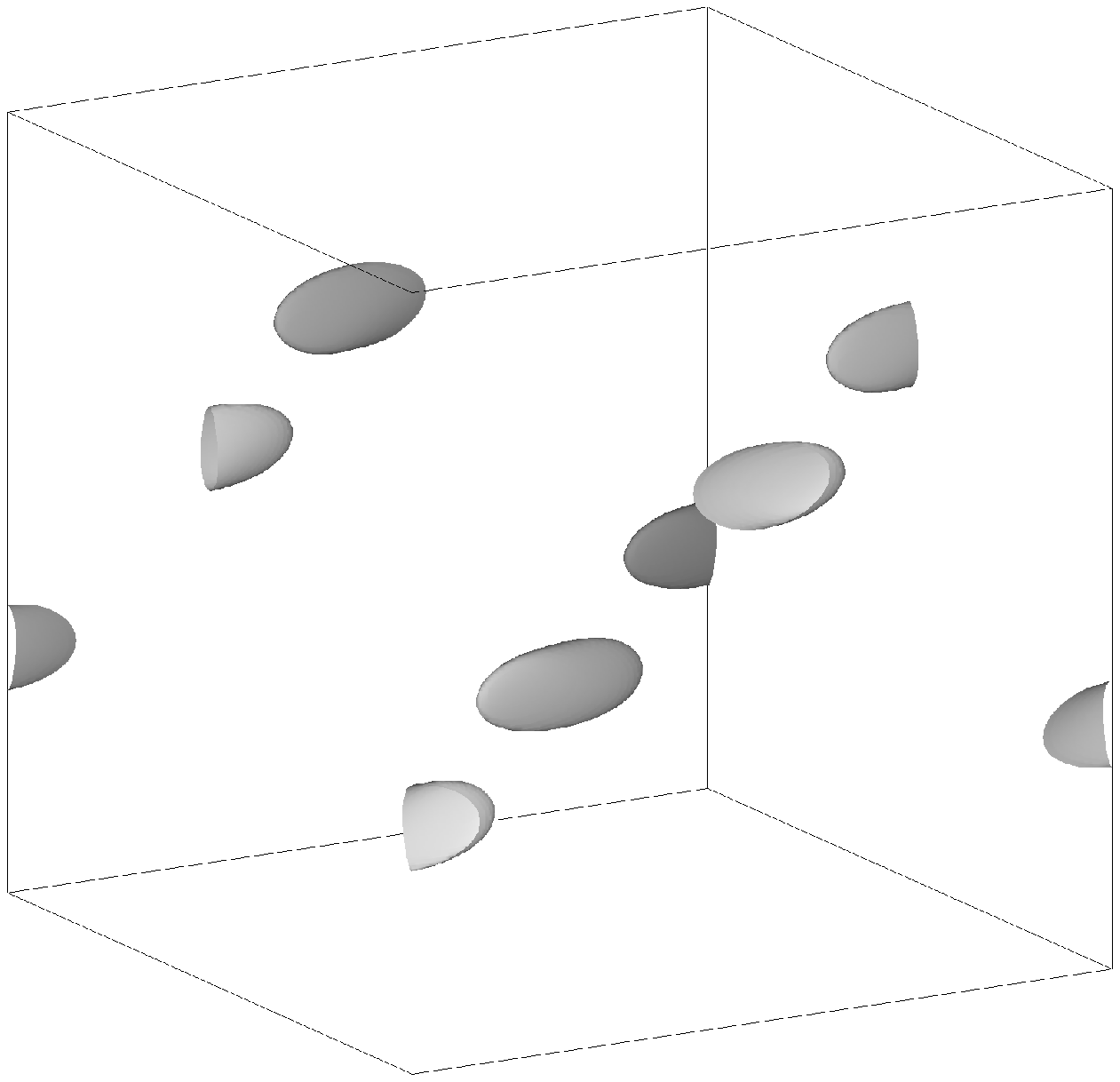,width=7cm,height=6cm,clip=}\hspace{1cm}
\psfig{file=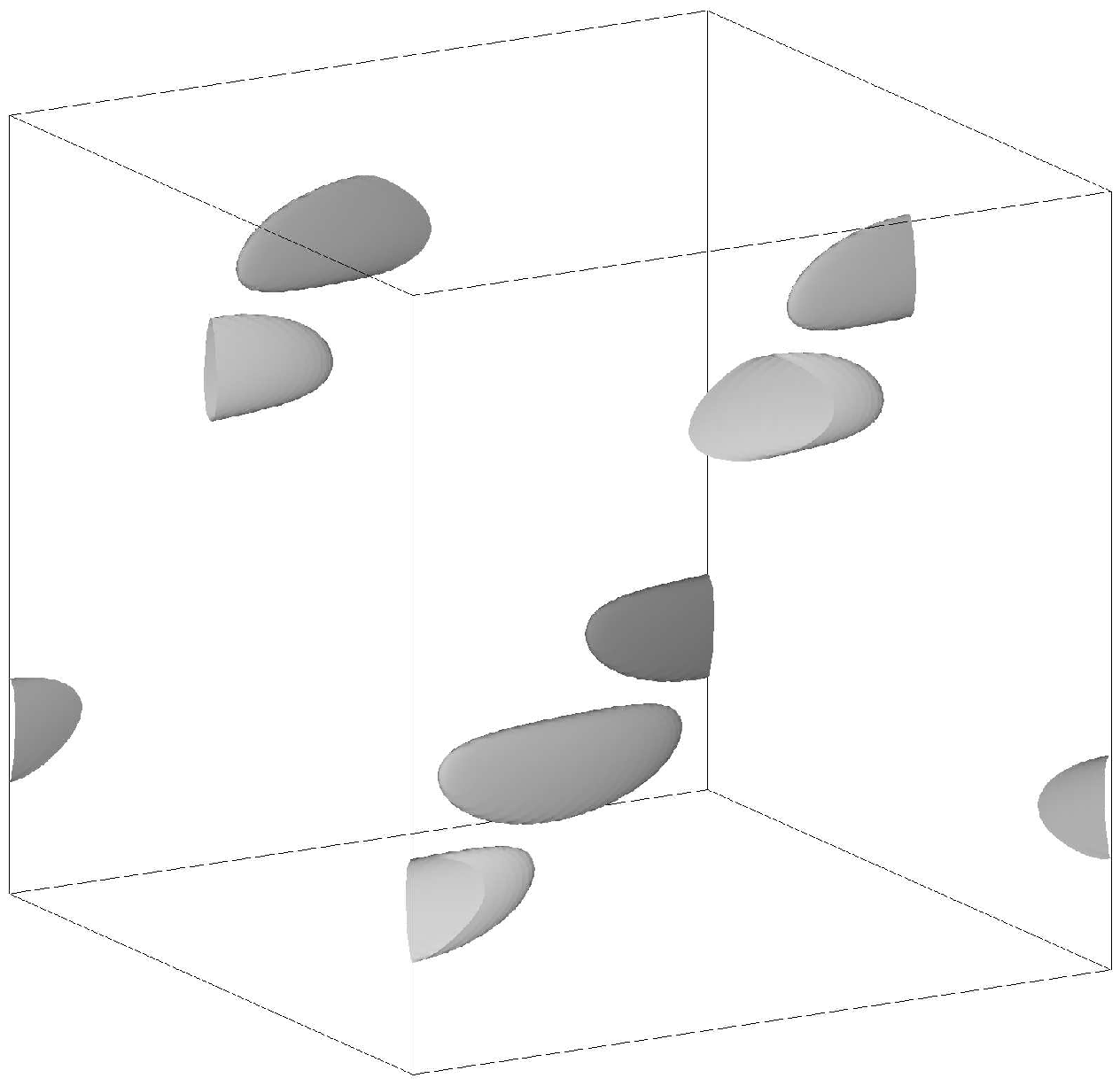,width=7cm,height=6cm,clip=}}

\vspace*{-4mm}
$t=0$\hspace{72mm}$t=T/8$

\vspace*{13mm}
\centerline{\psfig{file=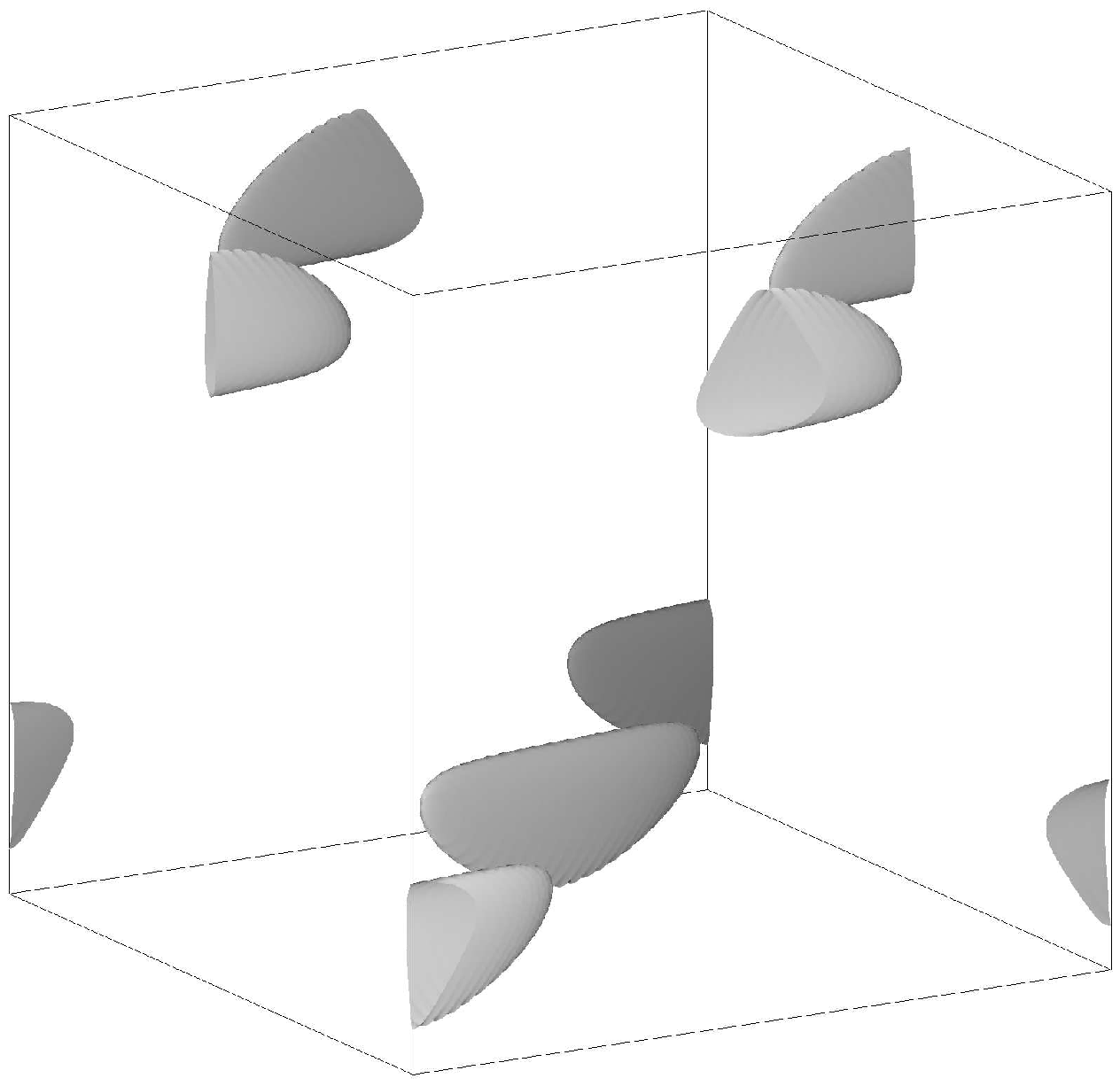,width=7cm,height=6cm,clip=}\hspace{1cm}
\psfig{file=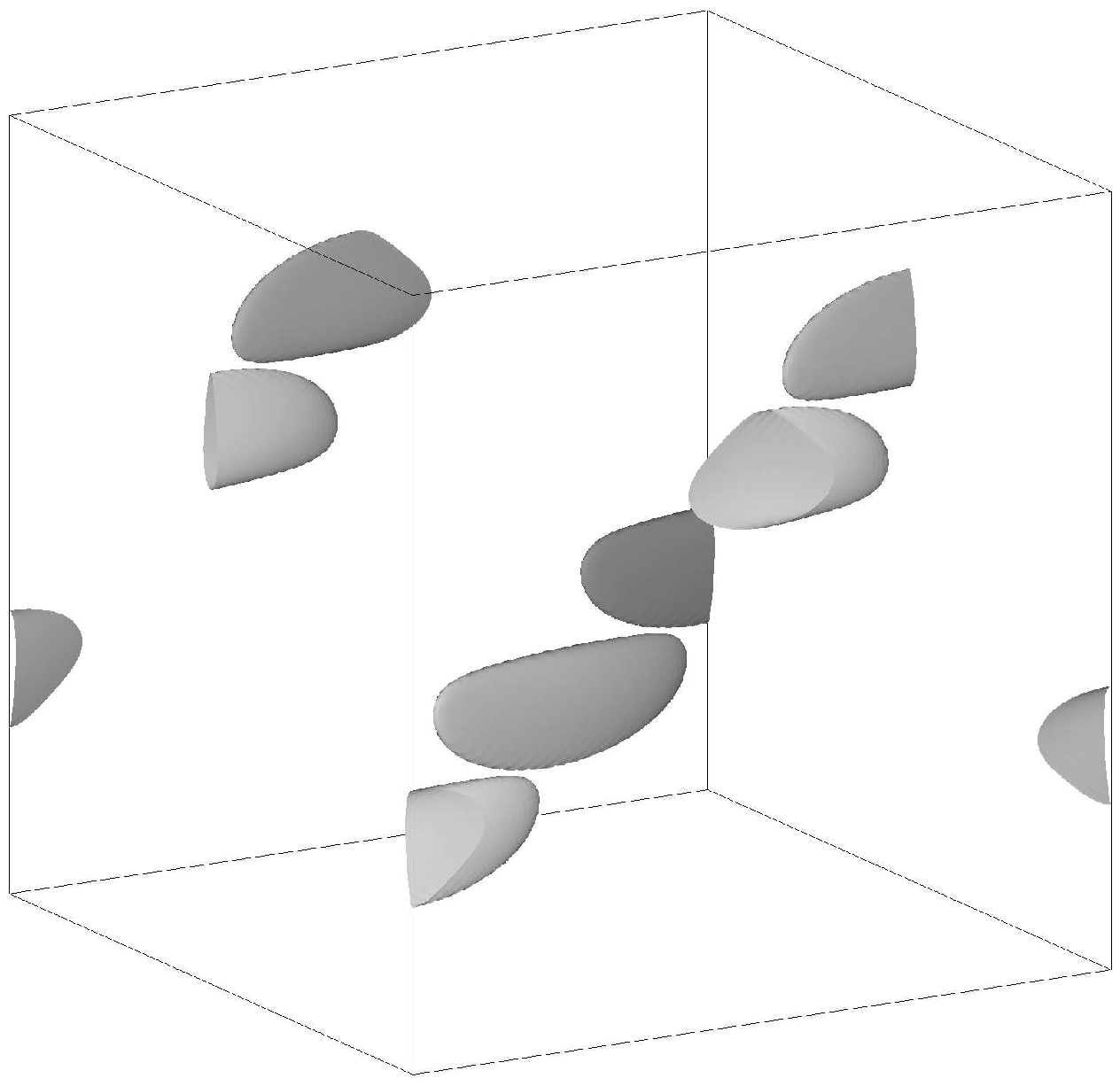,width=7cm,height=6cm,clip=}}

\vspace*{-4mm}
$t=T/4$\hspace{7cm}$t=3T/8$

\vspace*{13mm}
\centerline{\psfig{file=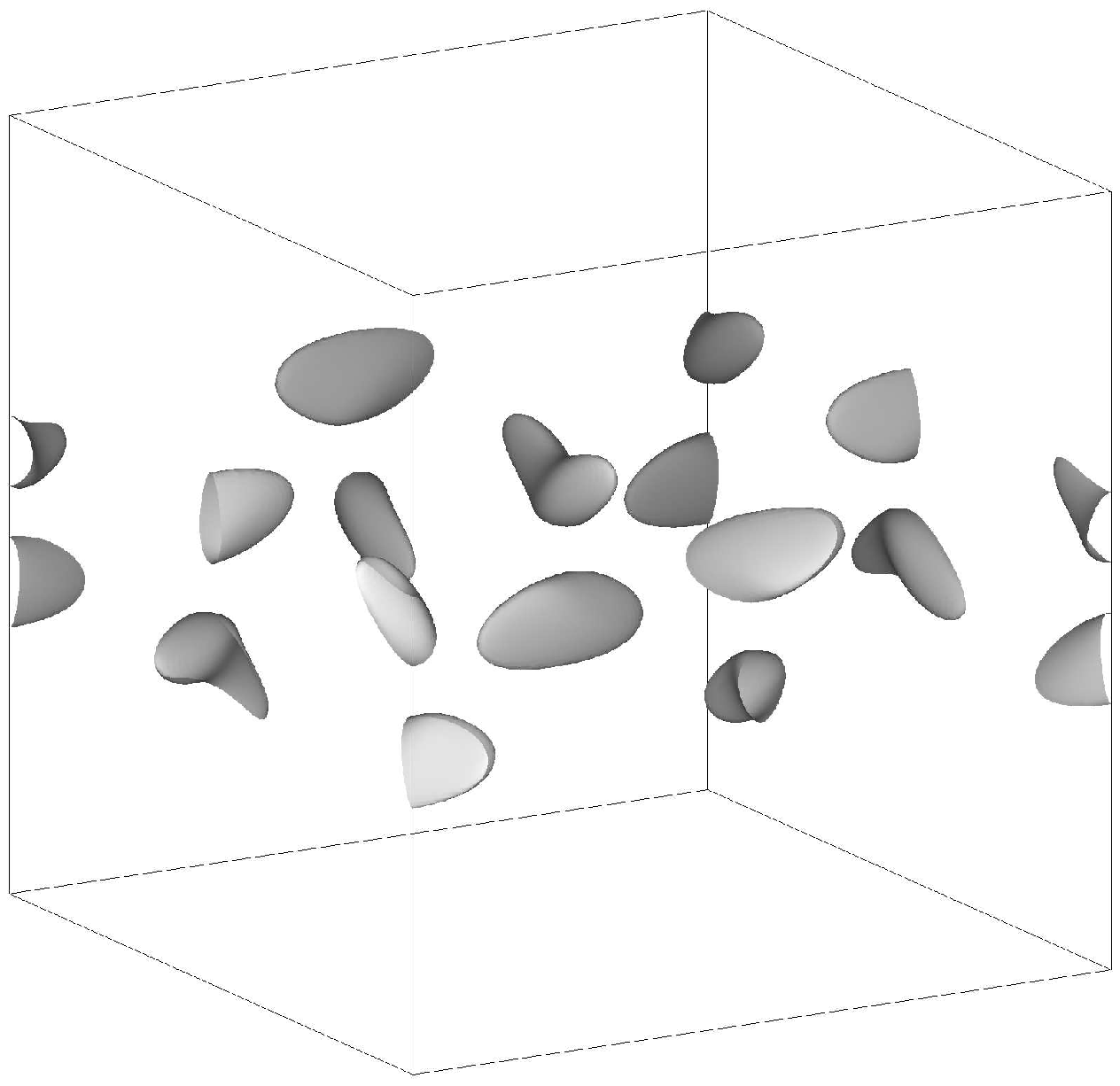,width=7cm,height=6cm,clip=}\hspace{1cm}
\psfig{file=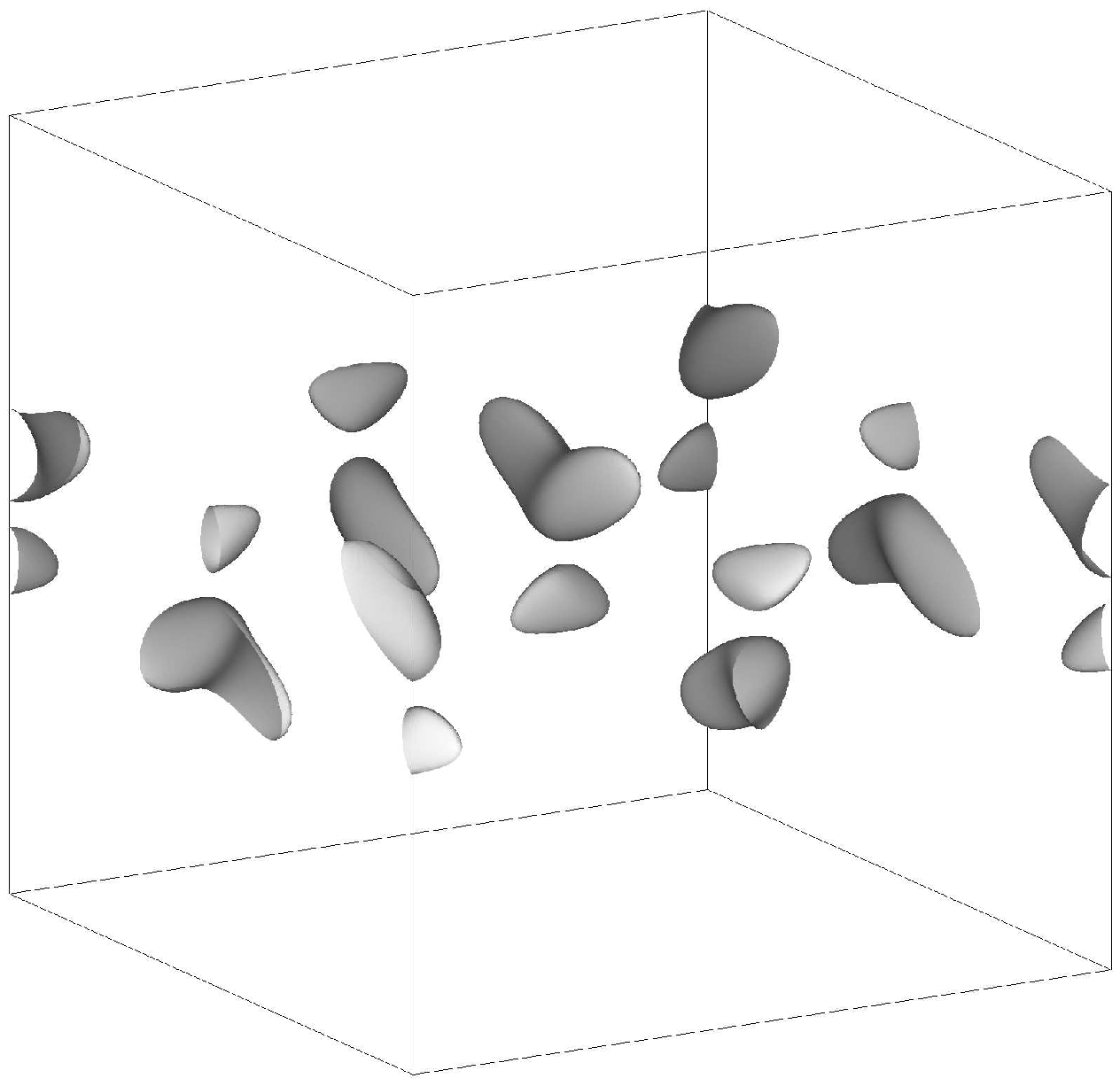,width=7cm,height=6cm,clip=}}

\vspace*{-4mm}
$t=T/2$\hspace{7cm}$t=5T/8$

\vspace*{8mm}

\caption{Isosurfaces of magnetic energy density $|{\bf h}|^2$ at the level
of 1/2 of the maximum for the time-periodic CHM regime for $R=5400$.
One periodicity cell is shown step 1/8 of the temporal period $T$.}
\label{f:eb5400}\end{figure}
\begin{figure}[t]
\centerline{\psfig{file=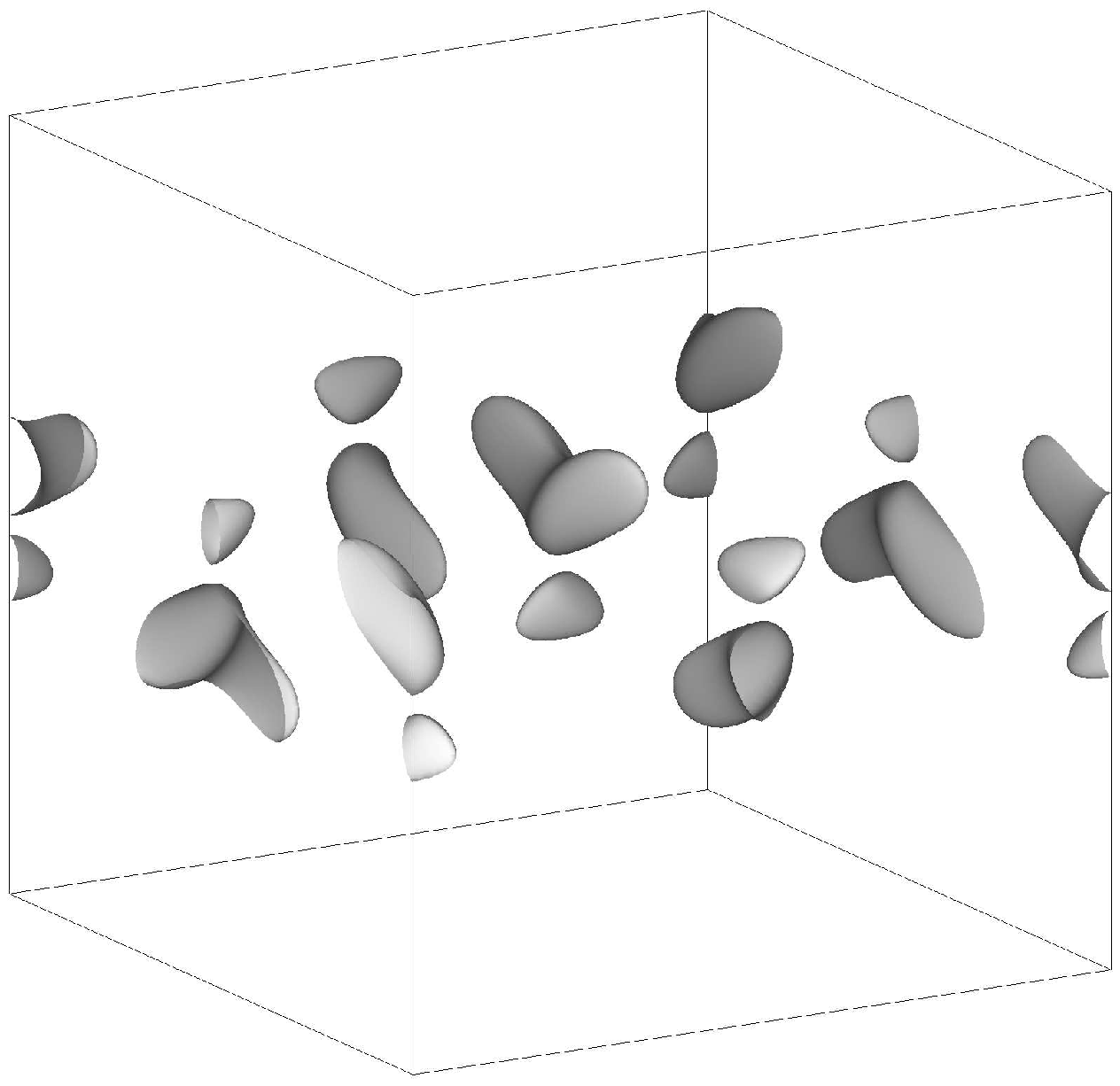,width=7cm,height=6cm,clip=}\hspace{1cm}
\psfig{file=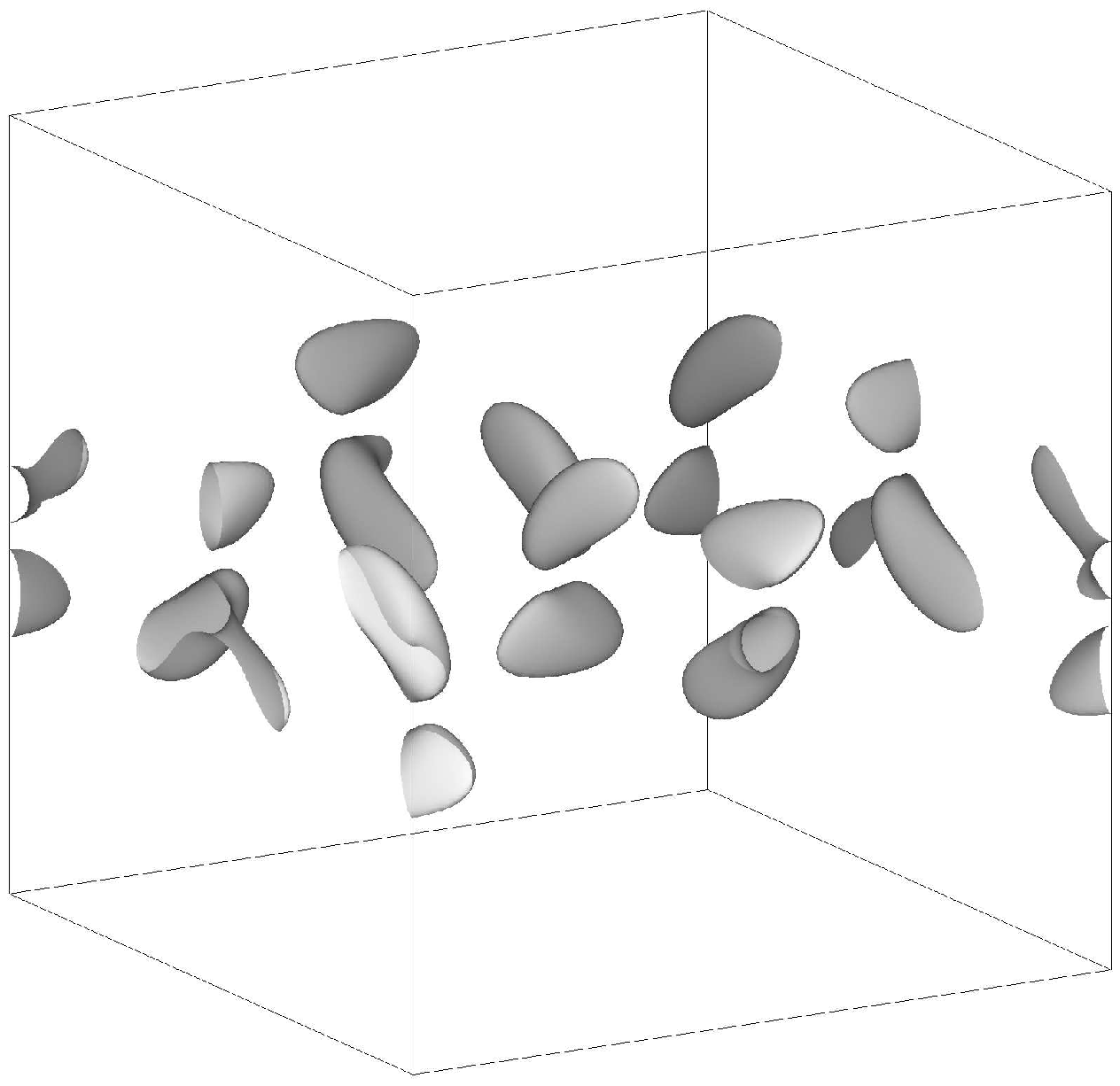,width=7cm,height=6cm,clip=}}

\vspace*{-4mm}
$t=3T/4$\hspace{7cm}$t=7T/8$

\vspace*{13mm}

\centerline{Figure \protect\ref{f:eb5400} (the end)}
\end{figure}

\begin{figure}[p]
\centerline{\psfig{file=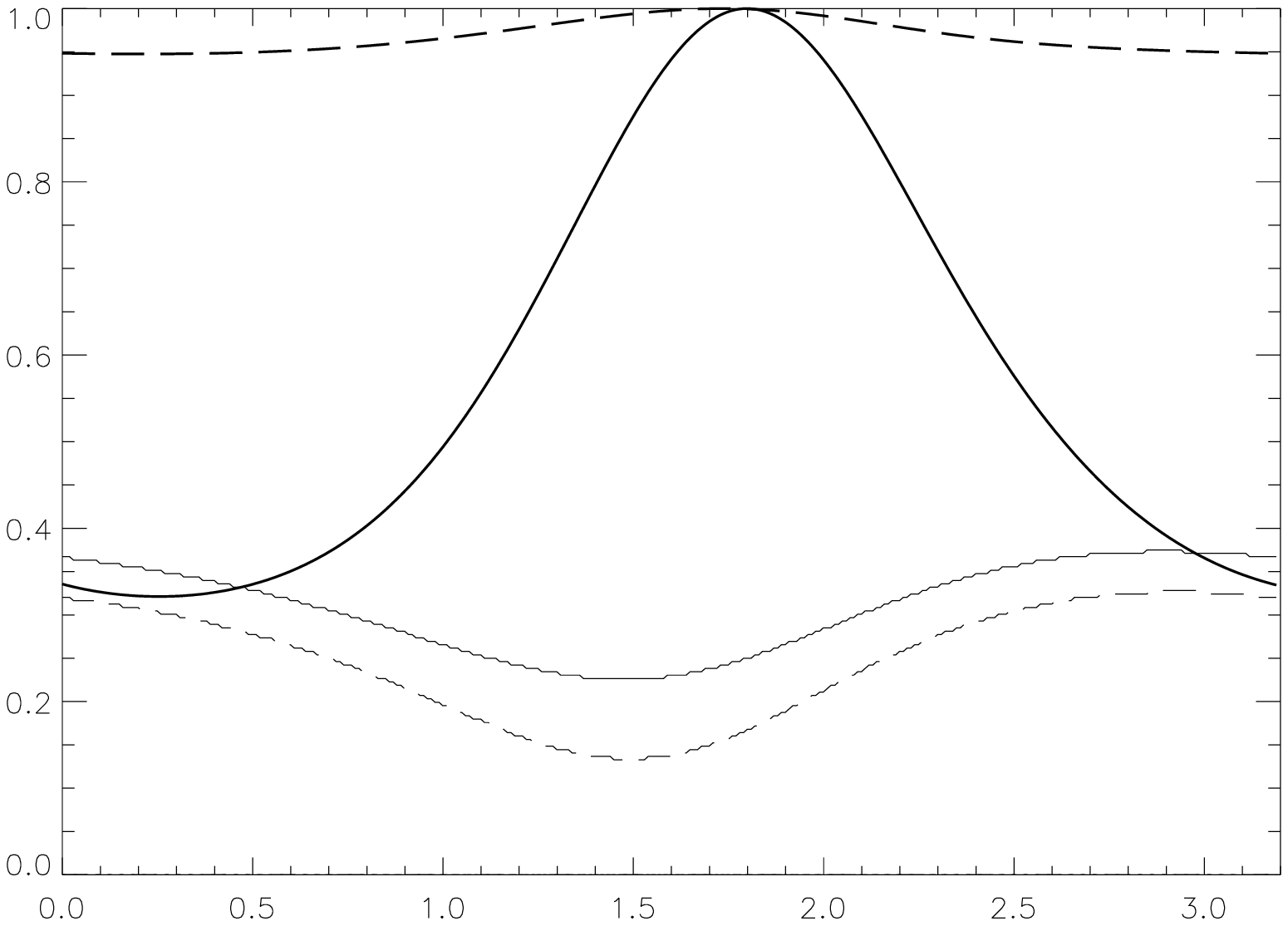,width=79mm,height=52mm,clip=}\hspace{2mm}
\psfig{file=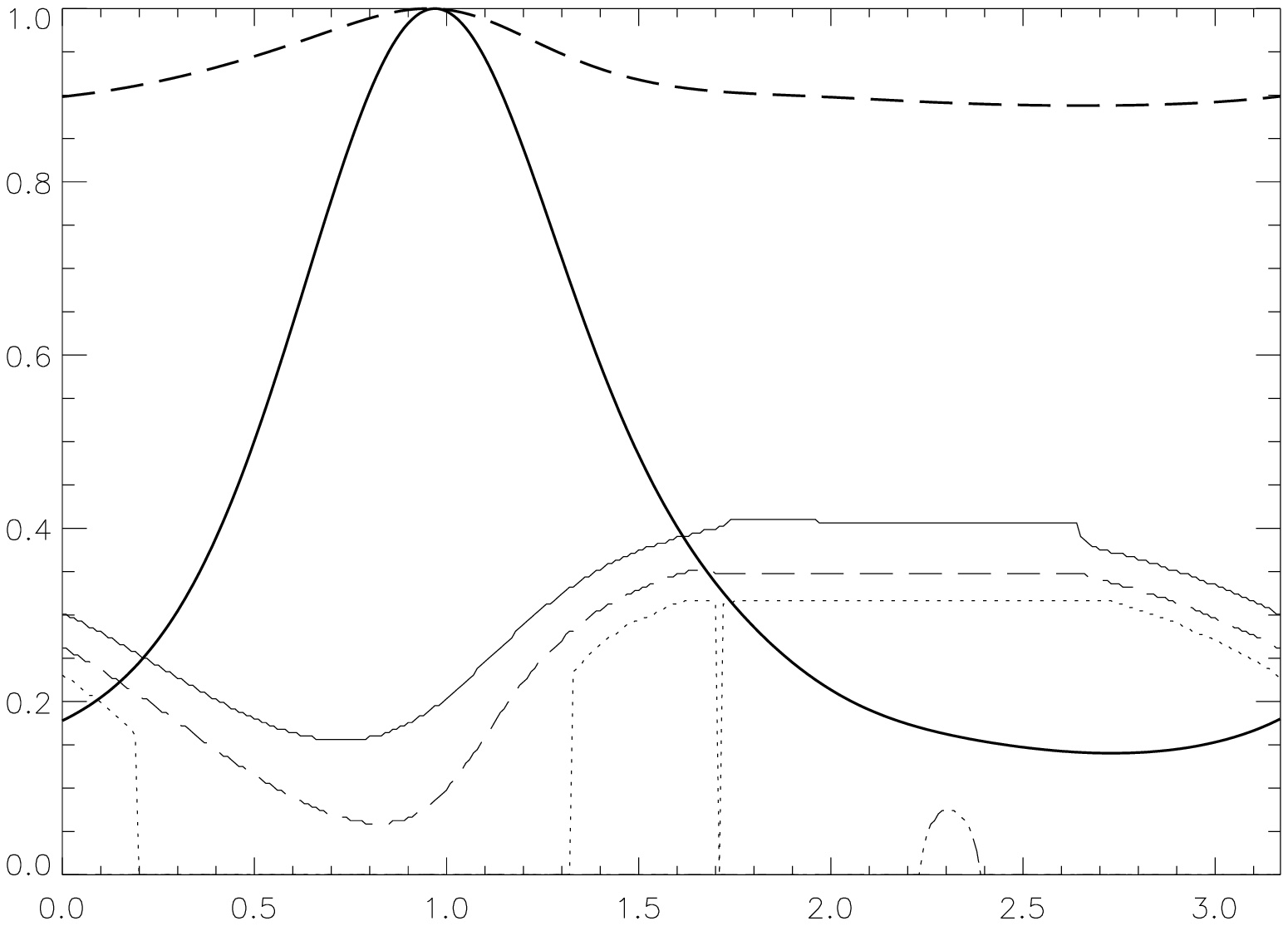,width=79mm,height=52mm,clip=}}

\vspace*{-49.5mm}
\hspace{5.5mm}$R=5350$

\vspace*{-2.5mm}
\hspace{14cm}$R=5400$

\vspace*{45mm}

\centerline{\psfig{file=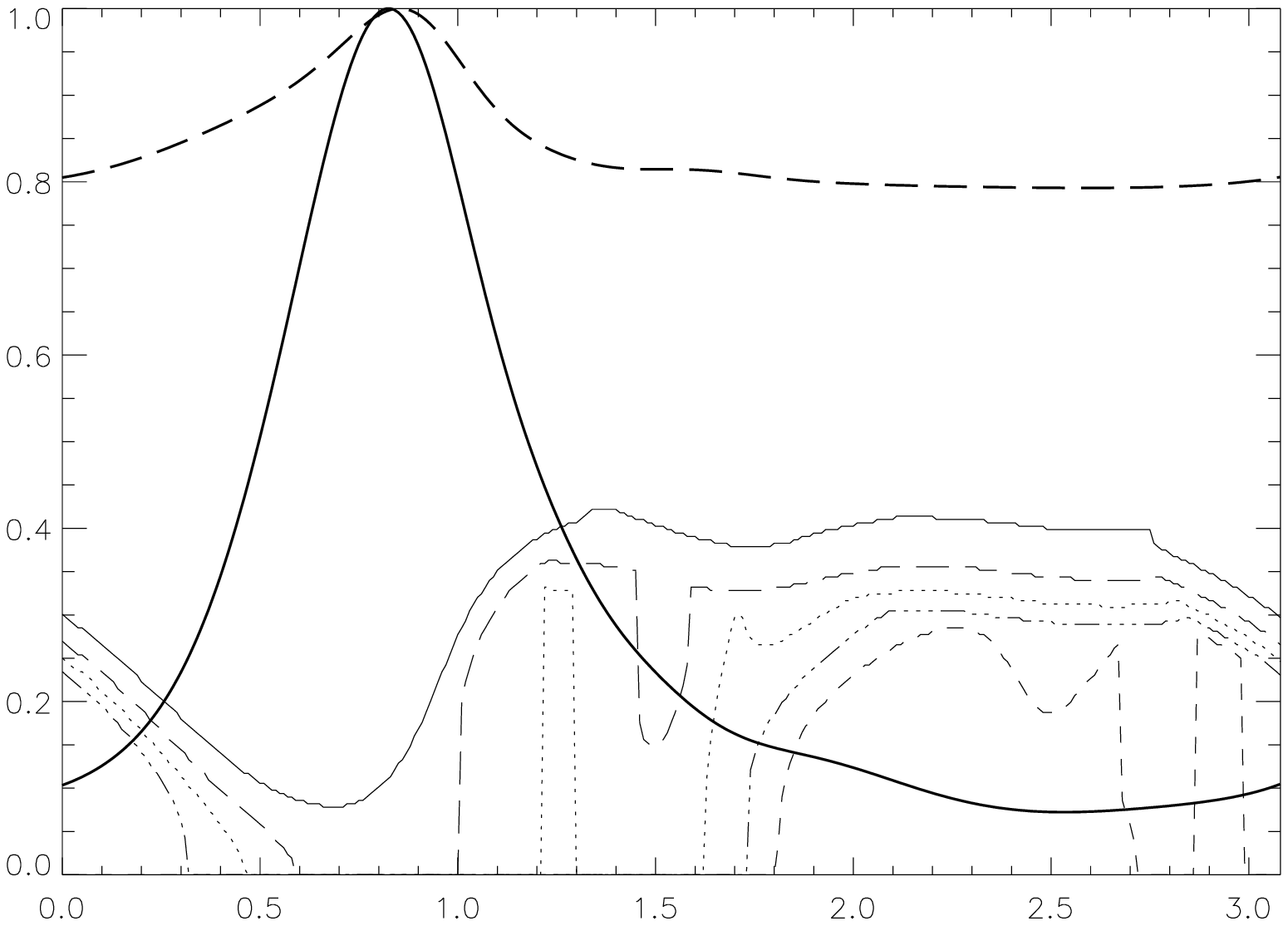,width=79mm,height=52mm,clip=}\hspace{2mm}
\psfig{file=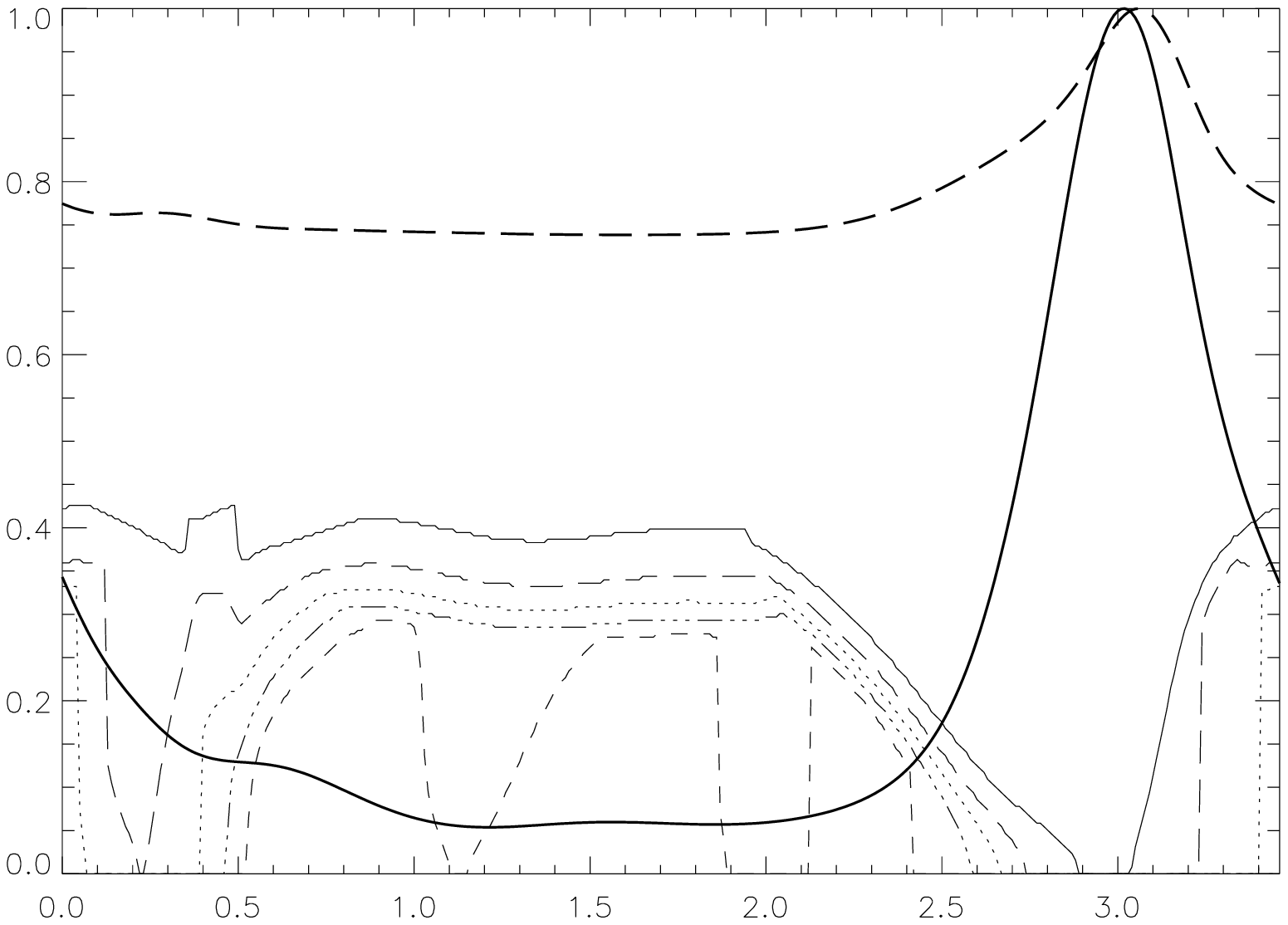,width=79mm,height=52mm,clip=}}

\vspace*{-52mm}
\hspace{58mm}$R=5500$\hspace{12.5mm}$R=5550$

\vspace*{51.5mm}

\centerline{\hspace{-.5mm}\psfig{file=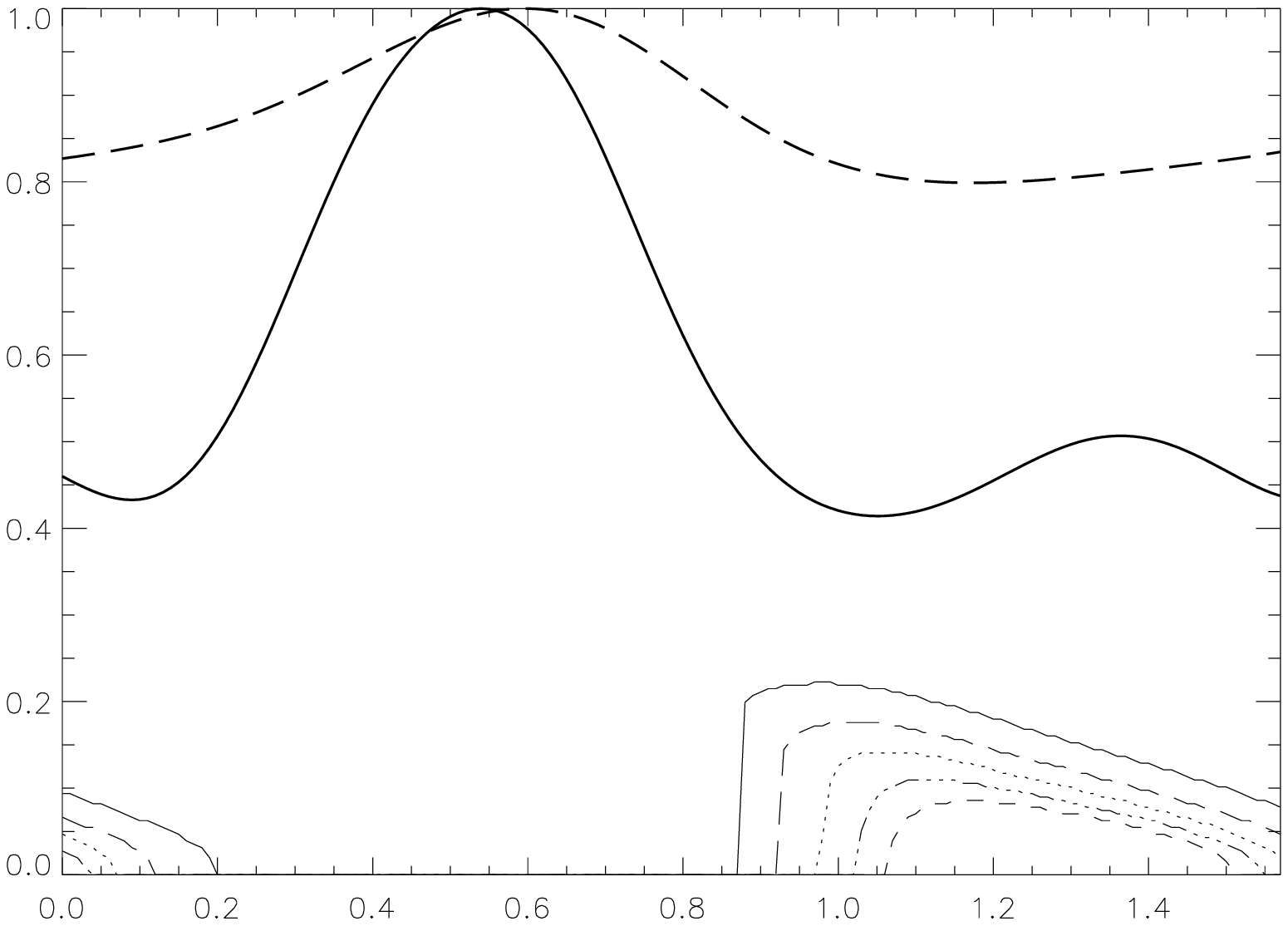,width=79mm,height=52mm,clip=}\hspace{82.5mm}}

\vspace*{-52mm}
\hspace{58mm}$R=5600$

\vspace*{44mm}

\caption{Distances (vertical axis) from sets, where magnetic energy density
is equal to or exceeds $e_{\max}(t)/16$ (thin short-dashed line),
$e_{\max}(t)/8$ (dash-and-dots line), $e_{\max}(t)/4$ (dotted line),
$e_{\max}(t)/2$ (thin long-dashed line) and ${15\over16}e_{\max}(t)$ (thin solid
line) to the closest boundaries of the fluid layer for $5350\le R\le 5600$
as a function of time (horizontal axis). Thick solid line: normalised total
magnetic energy $E^m(t)/\max_t E^m(t)$, thick dashed line: normalised total
kinetic energy $E^k(t)/\max_t E^k(t)$. One temporal period is shown.}
\label{f:dist}\end{figure}

\begin{figure}[t]
\centerline{\psfig{file=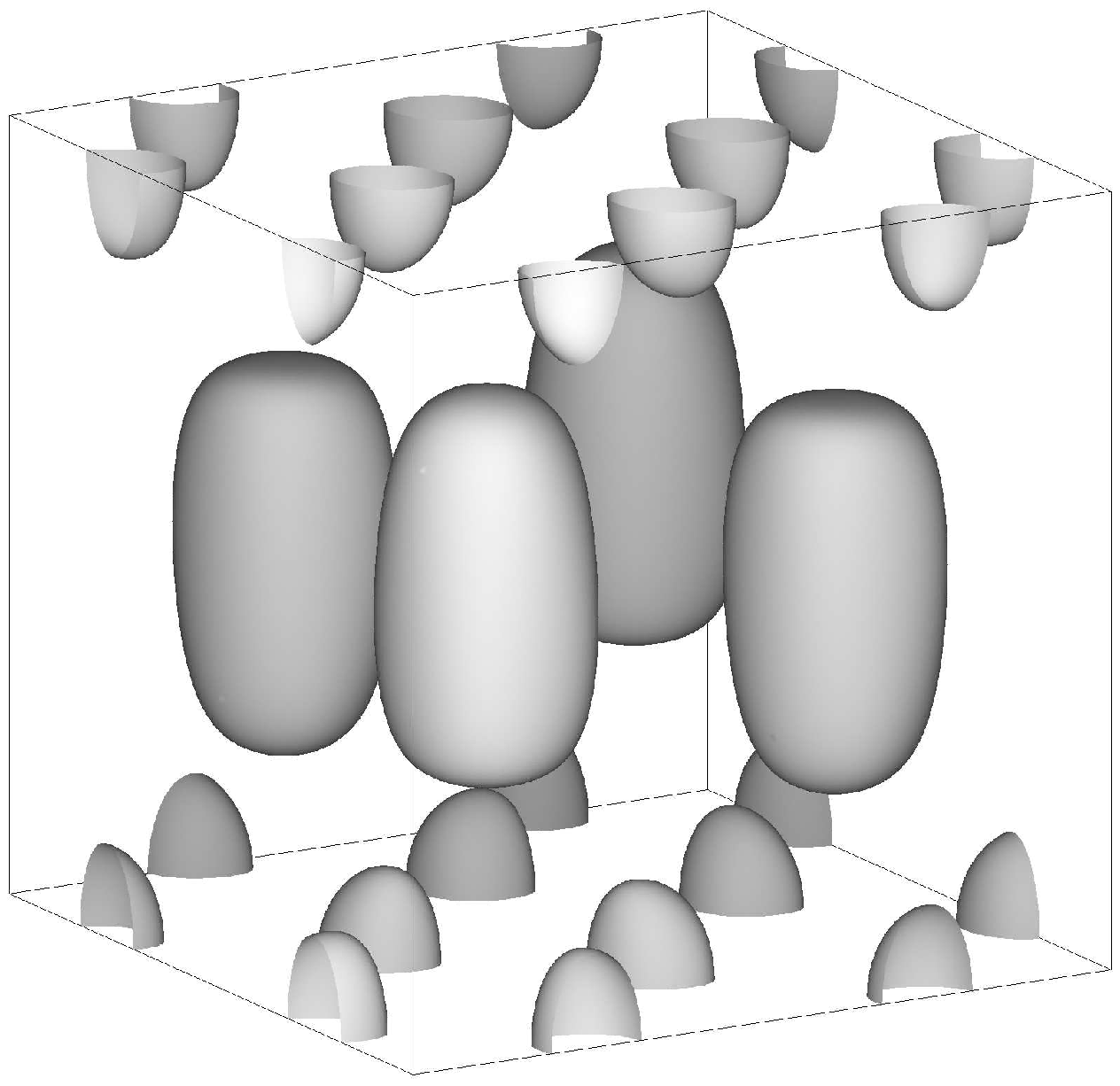,width=7cm,height=6cm,clip=}\hspace{1cm}
\psfig{file=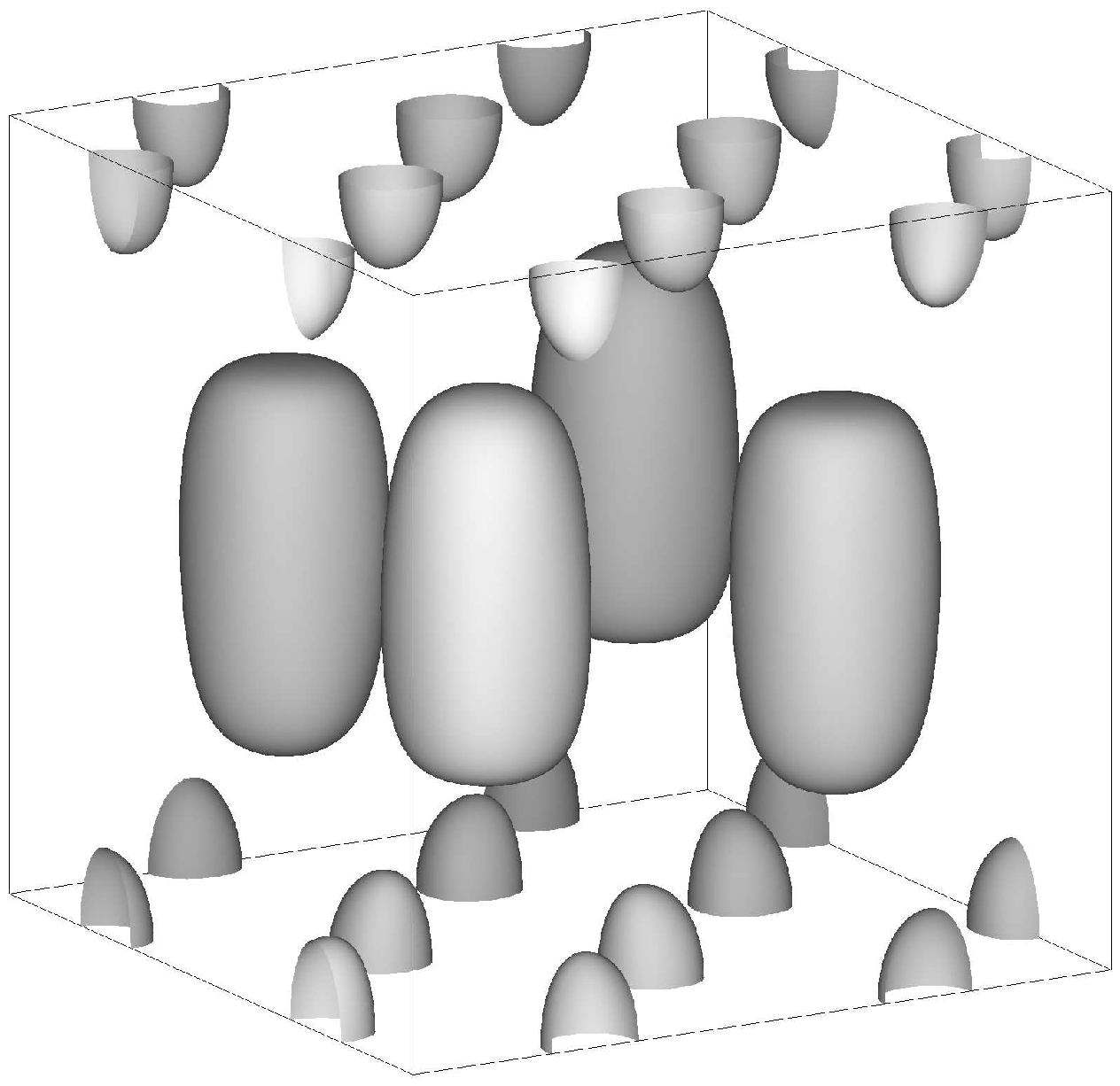,width=7cm,height=6cm,clip=}}

\vspace*{-7mm}
$t=0$\hspace{7cm}$t=T/2$

\caption{Isosurfaces of kinetic energy density $|{\bf v}|^2$ at the level
of a half of the maximum for the periodic CHM regime for $R=5400$.
One periodicity cell is shown step a half of the temporal period.}
\label{f:ev5400}

\bigskip
\centerline{\psfig{file=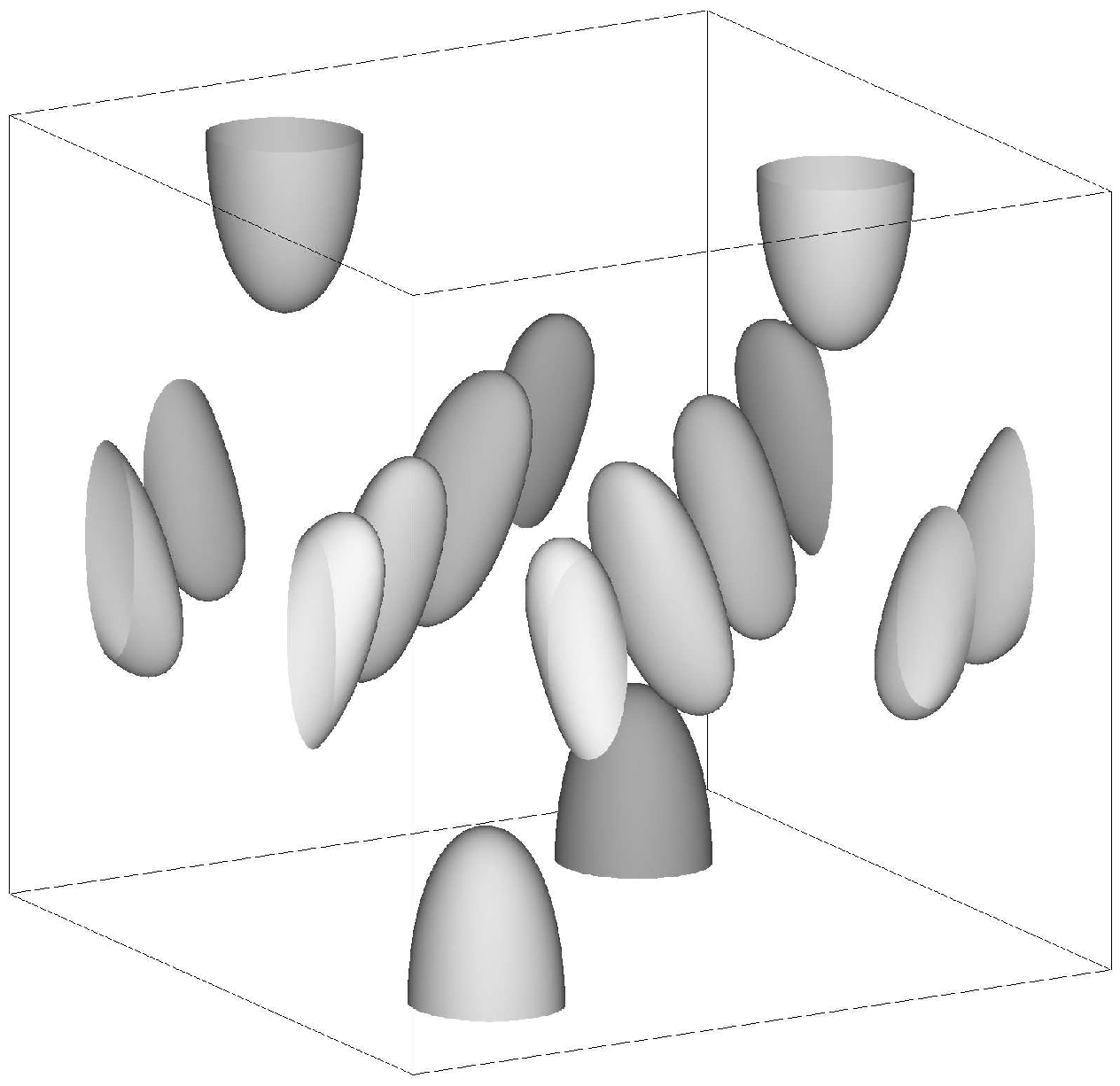,width=7cm,height=6cm,clip=}\hspace{1cm}
\psfig{file=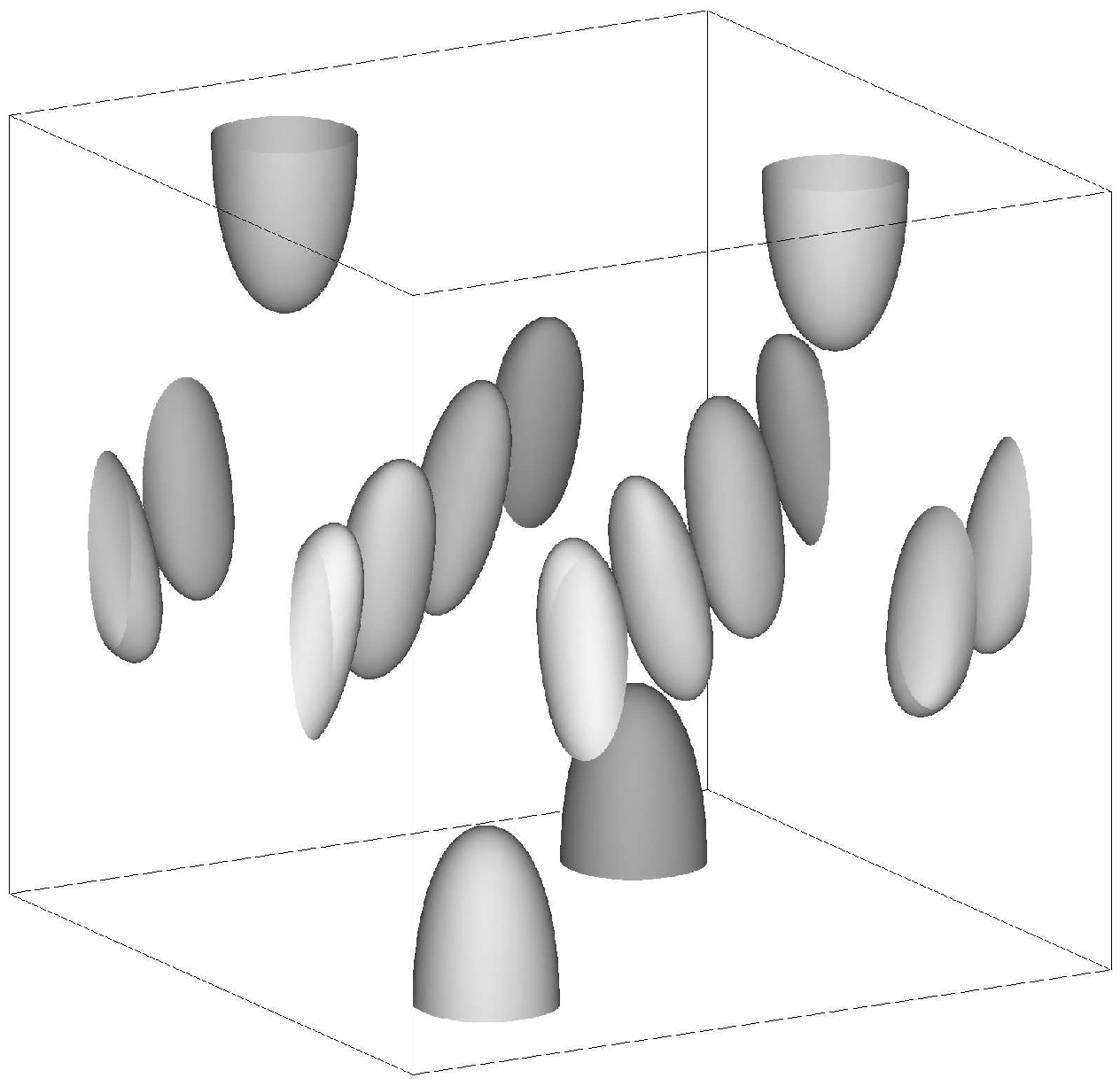,width=7cm,height=6cm,clip=}}

\vspace*{-7mm}
$t=0$\hspace{7cm}$t=T/2$

\caption{Isosurfaces of vorticity density $|\nabla\times{\bf v}|^2$
at the level of 2/3 of the maximum for the periodic CHM regime for $R=5400$.
One periodicity cell is shown step a half of the temporal period.}
\label{f:r5400}\end{figure}

The window of amagnetic convective rolls is followed by a sequence of periodic
regimes in the interval $5350\le R\le 5550$.
Periodic CHM regimes from the branch for $5350\le R\le 5550$ exhibit
an interesting feature: whilst for $5100\le R\le 5190$ the strongest magnetic
field is located on boundaries of the convective layer, in the former Rayleigh
number interval the maximum of magnetic field density is always
($5350\le R\le 5500$) or most of the time ($R=5550$) is in the interior of the
layer (see \xfg{f:eb5400}, \ref{f:dist}; segments of some plots are not visible
on \xfg{f:dist} for the times, where the distance vanishes). This is in contrast
with the behaviour found in all previous computations. For $0\le s\le1$, denote
by $U_s(t)$ the set of points in the layer, where at time $t$ magnetic energy
density $|{\bf h}|^2$ is equal to or exceeds the value $se_{\max}(t)$,
$e_{\max}(t)$ being the maximum of magnetic energy density spatial distribution
at time $t$. For $R=5350$ and 5400, the set $U_{15/16}(t)$ is always at
a distance of at least 0.226 and 0.156 of the layer width away from
the boundaries, and $U_{1/2}(t)$ is not less than 0.133 and 0.057 of the width
away, respectively. Consequently, these CHM regimes may be not very sensitive
to the boundary conditions for magnetic field. In all simulated CHM regimes
the distance is the smallest (in time) near the points of maxima of the total
kinetic and magnetic energies. Concentration of magnetic field in the interior
of the layer can be related to the fact, that the regions of the most vigorous
flow motions are inside it (\xfg{f:ev5400}, \ref{f:r5400}). Since isosurfaces
shown on these figures only weakly depend on time, they are presented step
a half of the temporal period, when the difference in their shape is extreme.
(Vorticity admits the maximum on the boundaries of the layer; both
the flow and vorticity are strong in some regions close to the boundaries,
but the size of these regions is comparatively small, which may explain
why they do not influence magnetic field generation significantly.)

The branch existing for $5350\le R\le 5550$ apparently disappears between
$R=5550$ and 5600. For $R=5600$ the regime is a periodic orbit of temporal
frequency more than twice higher (see the table). It has a different symmetry
group: it is the only periodic orbit that we have found, where the flow
is not symmetric about a vertical axis. Between $R=5600$ and 5700 the periodic
regime changes once again. At $R=5800$ the CHM regime is a chaotic attractor
of heteroclinic nature. Like for $R=5210$, the sample trajectory enjoys
excursions between a steady state and a periodic orbit (in this description
we do not distinguish states related by discrete symmetries and translation
in horizontal directions), but apart from this the
patterns of behaviour of the two chaotic attractors are completely different.
The periodic orbit emerges from the one for $R=5700$ in a period-doubling
bifurcation; in particular, it has all the parity-related symmetries
of the latter. This periodic orbit is stable at $R=5750$, but becomes unstable
at $R=5800$. The steady state is amagnetic rolls aligned with the diagonal
of the periodicity cell and invariant with respect to translations along
the axis; therefore, it also possesses all the symmetries of the periodic
orbit at $R=5700$. However, all the symmetries are broken during the exponential
transition processes when the trajectory moves between the two states,
except for the composition of translation in the $x_2-$direction and
reflection about the mid-plane $\sigma_3$. Both states are only marginally
unstable. The rolls are apparently capable of kinematic generation of magnetic
field, but while the trajectory approaches them, magnetic energy falls off
almost to zero (\xfg{f:e5800}).

\begin{figure}[p]
\centerline{\raisebox{1cm}{(a)}~~~\psfig{file=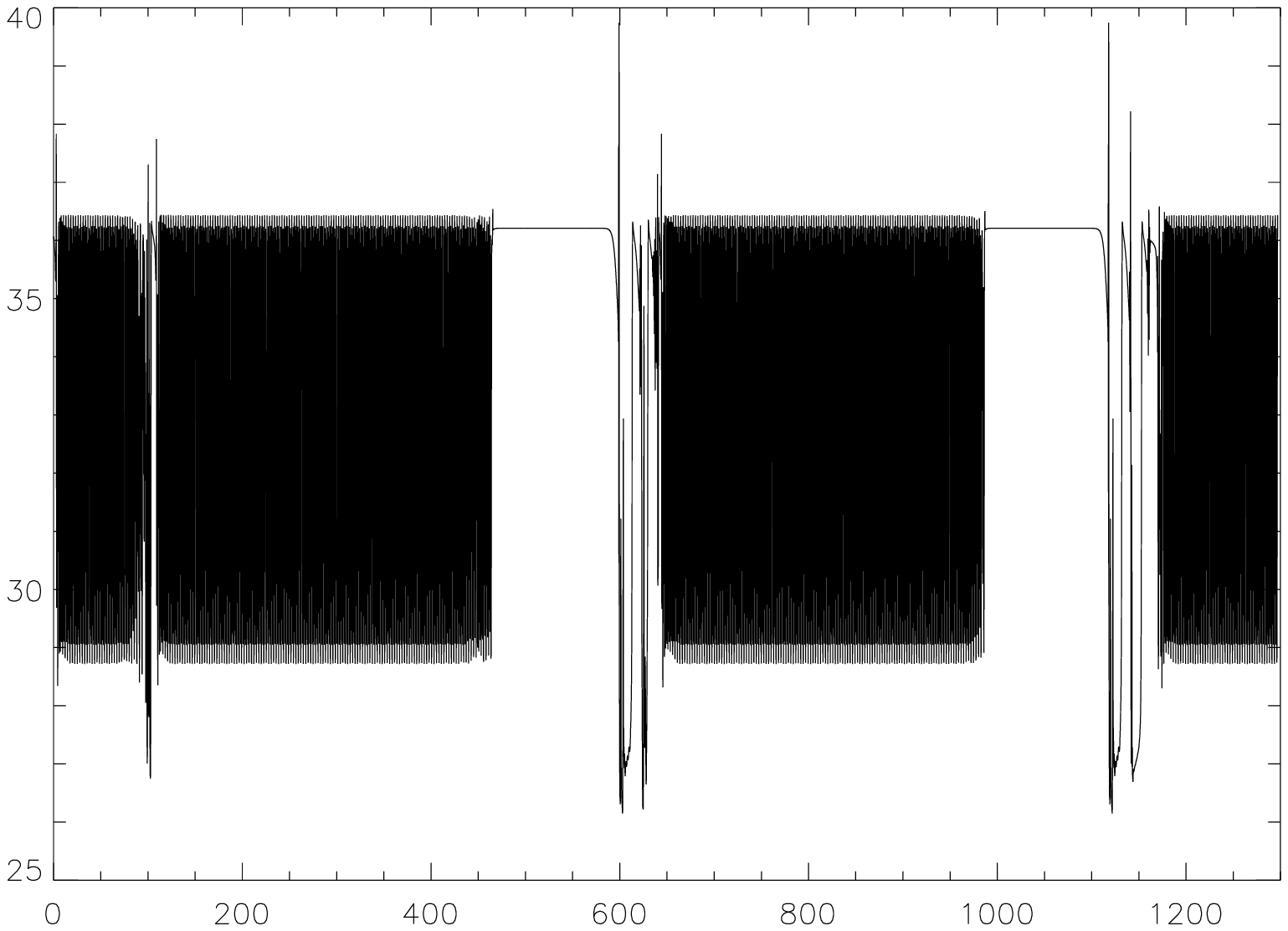,width=15cm,height=9cm,clip=}}
\vspace*{15mm}

\centerline{\raisebox{1cm}{(b)}~~~\psfig{file=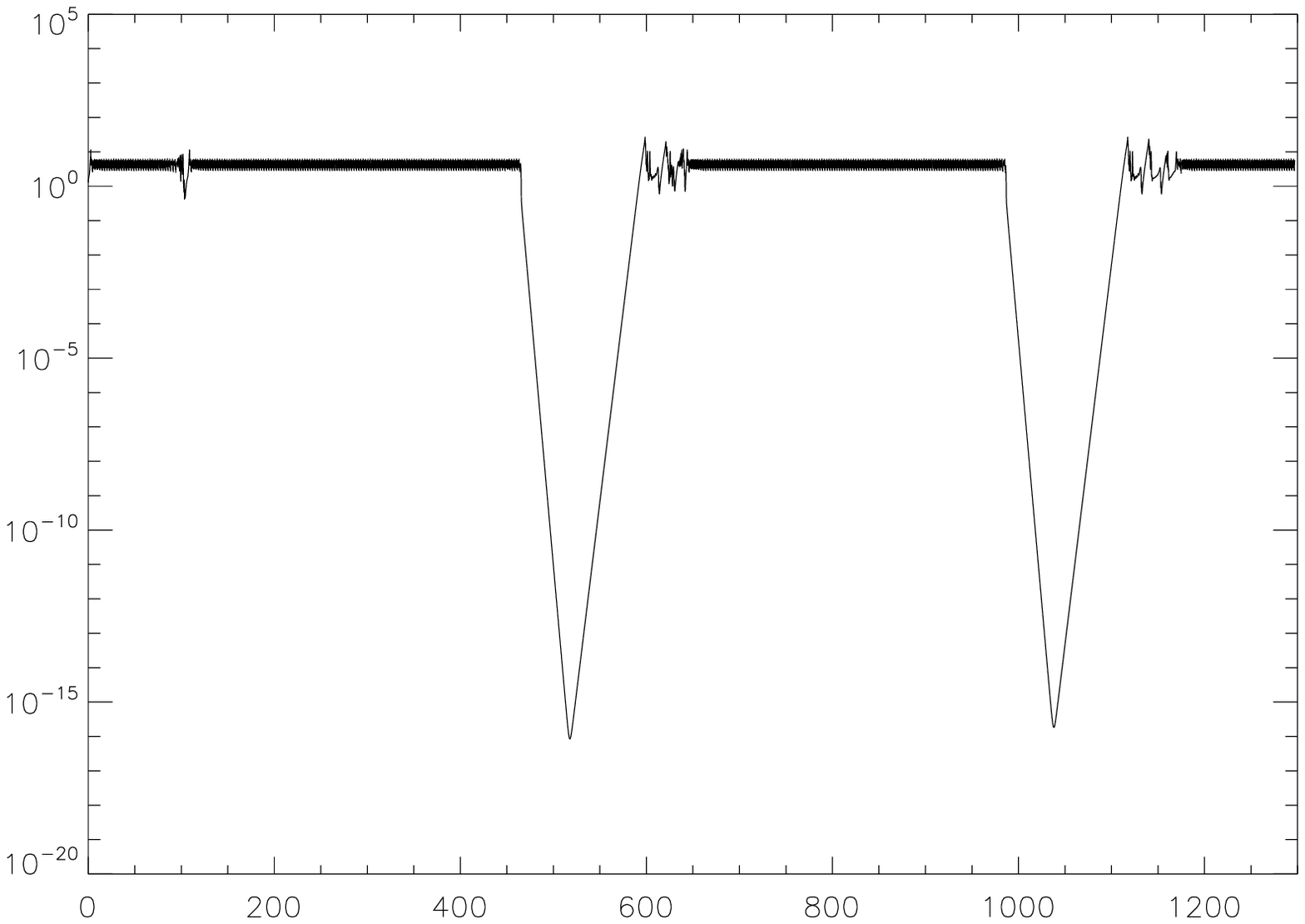,width=15cm,height=9cm,clip=}}
\caption{Total kinetic (a) and magnetic (b) energies (vertical axes) as
functions of time (horizontal axis) for the chaotic CHM regime for $R=5800$.
Magnetic energy is shown in logarithmic scale.}
\label{f:e5800}\end{figure}

\section{Conclusion}

We have observed a number of steady and periodic CHM attractors,
possessing the symmetry about the vertical axis, and several regimes,
where this symmetry is broken partially (the magnetic field becoming
antisymmetric) or completely. The amplitude equations derived in \cite{Zh09}
are applicable for examination of the weakly non-linear stability
of the symmetric CHM regimes presented here. It remains to be investigated
how disappears the branch of periodic orbits, existing for
$5350\le R\le 5550$. If this is a symmetry-breaking bifurcation, the system
of amplitude equations for free CHM convection, analogous to the system
of mean-field equations derived in \cite{Zh08} for a branch of regimes
emerging in a Hopf bifurcation in forced CHM convection, might be applicable
for examination of weakly nonlinear large-scale stability of the periodic
regimes constituting this branch near this bifurcation.

We have found two other types of notable convective
magnetic dynamos. One of the regimes is periodic orbits for Rayleigh numbers
from the interval $5350\le R\le 5500$: it is apparently the first known
example of CHM regimes, where the strongest generated magnetic field always
resides in the interior of the fluid layer. The second is the heteroclinic
chaotic regime for $R=5800$, in which the evolution consists of phases,
where the regime is in the vicinity either of a periodic orbit, or a steady
state -- amagnetic rolls. Dynamics of this attractor can be studied
employing expanded central manifold reduction \cite{P06a,P06b}.

\mi{\bf Acknowledgments}

\mi
I have benefited from discussions with J.-F.~Pinton. Comments of an anonymous
Referee have helped to improve the paper.
Part of this research was carried out during the visit of the author to the
School of Engineering, Computer Science and Mathematics, University of Exeter,
UK, in January -- April 2008. I am grateful to the Royal Society for their
financial support. Computations have been carried out on the computer
``M\'esocentre SIGAMM (Simulations Interactives et Visualisation en
G\'eophysique, Astronomie, Math\'ematique et M\'ecanique)'' hosted by
Observatoire de la C\^ote d'Azur, France. My research work at the Observatoire
de la C\^ote d'Azur in the autumns of 2007 and 2008 was supported by the French
Ministry of Education. I was partially financed by the grants ANR-07-BLAN-0235
OTARIE from Agence nationale de la recherche, France, and
07-01-92217-CNRSL{\Large\_}a from the Russian foundation for basic research.

\mi
\end{document}